\documentclass{article}
\PassOptionsToPackage{numbers,compress}{natbib}

\usepackage[final]{neurips_2025}

\usepackage[utf8]{inputenc} 
\usepackage[T1]{fontenc}    
\usepackage{url}            
\usepackage{booktabs}       
\usepackage{amsfonts}       
\usepackage{nicefrac}       
\usepackage{microtype}      
\usepackage{xcolor}         
\usepackage{graphicx}
\usepackage{enumitem}
\usepackage{amsmath}
\usepackage{algorithm}
\usepackage{algorithmic}
\usepackage{multirow}
\usepackage{makecell}
\usepackage[usenames,dvipsnames]{xcolor}
\usepackage{hyperref}  
\usepackage{wrapfig}

\title{Semantic Retrieval Augmented Contrastive Learning for Sequential Recommendation}

\author{%
  Ziqiang Cui\textsuperscript{1\thanks{Equal contribution.}},  
  Yunpeng Weng\textsuperscript{2 3\footnotemark[1]}, 
  Xing Tang\textsuperscript{4\footnotemark[2]}, 
  Xiaokun Zhang\textsuperscript{1},
  Shiwei Li\textsuperscript{2},
  Peiyang Liu\textsuperscript{5}, \\
  \textbf{Bowei He\textsuperscript{1},
  Dugang Liu\textsuperscript{6},
  Weihong Luo\textsuperscript{3},
  Xiuqiang He\textsuperscript{4},
  Chen Ma\textsuperscript{1\thanks{Corresponding authors.}},
      }
  \\
  \footnotesize
  \textsuperscript{1} City University of Hong Kong \quad
    \footnotesize  
  \textsuperscript{2} Huazhong University of Science and Technology \\
  \footnotesize
  \textsuperscript{3} Tencent \quad
  \footnotesize
  \textsuperscript{4} Shenzhen Technology University \quad
\footnotesize
  \textsuperscript{5} Peking University \quad
  \footnotesize  
  \textsuperscript{6} Shenzhen University \\
  \footnotesize
  ziqiang.cui@my.cityu.edu.hk, \{wengyp, lishiwei\}@hust.edu.cn, 
  xing.tang@hotmail.com, \\
  \footnotesize
  \{dawnkun1993, dugang.ldg\}@gmail.com, liupeiyang@pku.edu.cn,  boweihe2-c@my.cityu.edu.hk, \\ 
    \footnotesize
lobby66@163.com, 
  hexiuqiang@sztu.edu.cn,
  chenma@cityu.edu.hk
}

\begin{document}

\maketitle

\begin{abstract} 
Contrastive learning has shown effectiveness in improving sequential recommendation models. However, existing methods still face challenges in generating high-quality contrastive pairs: they either rely on random perturbations that corrupt user preference patterns or depend on sparse collaborative data that generates unreliable contrastive pairs. Furthermore, existing approaches typically require predefined selection rules that impose strong assumptions, limiting the model's ability to autonomously learn optimal contrastive pairs. To address these limitations, we propose a novel approach named Semantic Retrieval Augmented Contrastive Learning (SRA-CL). SRA-CL leverages the semantic understanding and reasoning capabilities of LLMs to generate expressive embeddings that capture both user preferences and item characteristics. These semantic embeddings enable the construction of candidate pools for inter-user and intra-user contrastive learning through semantic-based retrieval. To further enhance the quality of the contrastive samples, we introduce a learnable sample synthesizer that optimizes the contrastive sample generation process during model training. SRA-CL adopts a plug-and-play design, enabling seamless integration with existing sequential recommendation architectures. Extensive experiments on four public datasets demonstrate the effectiveness and model-agnostic nature of our approach. 
Our code is available at
\textcolor{blue}{\url{https://github.com/ziqiangcui/SRA-CL}}
\end{abstract}

\section{Introduction} \label{intro_sec}
Sequential recommendation aims to model user preferences based on historical behavior sequences, a task of significant value for online platforms like YouTube and Amazon. However, accurate preference modeling faces a fundamental challenge: data sparsity, as most users have only limited interaction records and most items receive little attention. To address this issue, numerous self-supervised learning techniques \cite{yu2023self,zhou2020s3,xie2022contrastive} have been proposed, leveraging auxiliary tasks to improve data utilization efficiency. Among these, contrastive learning has emerged as a predominant approach due to its conceptual simplicity and proven effectiveness \cite{xie2022contrastive,qiu2022contrastive,chen2022intent,qin2023meta,zhou2023equivariant}. 
Typically, it constructs positive sample pairs from the data and maximizes their agreement in the representation space \cite{chen2020simple}.

\begin{figure}[t]
\setlength{\belowcaptionskip}{-5mm} 
  \centering
\includegraphics[width=0.95\textwidth]{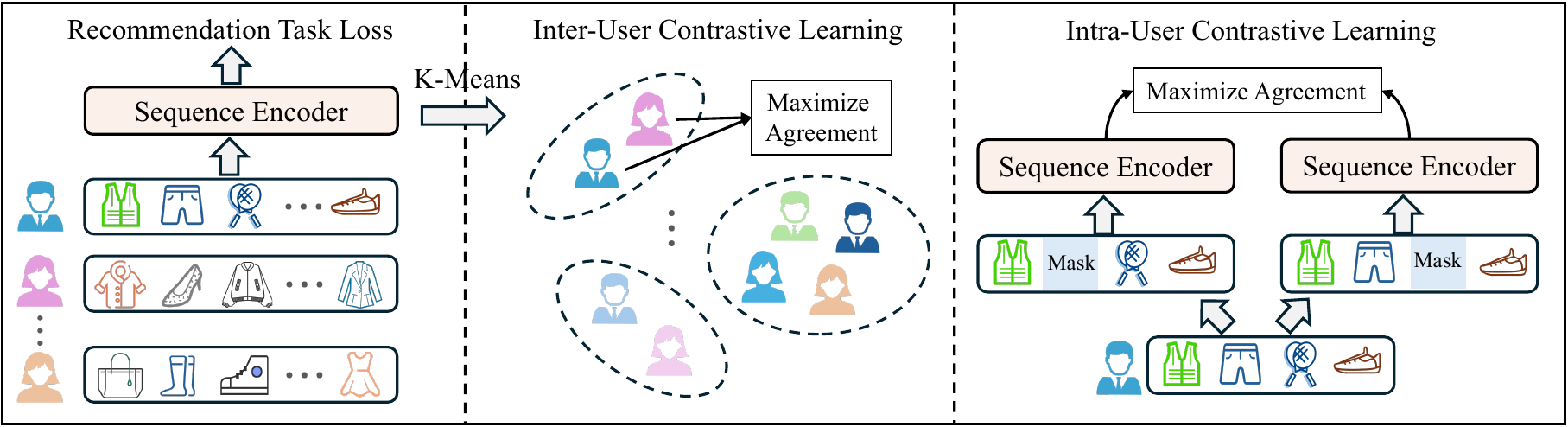}
  \caption{Illustration of existing contrastive learning methods in sequential recommendation, categorized into two main types: (1) inter-user contrastive learning and (2) intra-user contrastive learning.} 
   \label{intro_pdf}
\end{figure}

As illustrated in Figure \ref{intro_pdf}, existing contrastive learning approaches for sequential recommendation can be broadly classified into two categories: (1) inter-user contrastive learning, which contrasts sequences from different users, and (2) intra-user contrastive learning, which contrasts different augmented views of a single user's sequence. In the inter-user paradigm, user sequence representations are clustered using K-means, and users within the same cluster are treated as positive samples for each other~\cite{chen2022intent,li2023multi,qin2024intent}. In the intra-user paradigm, perturbations are applied to a user's sequence to generate augmented views, and the similarity between these views is maximized~\cite{xie2022contrastive,liu2021contrastive,qiu2022contrastive,qin2023meta}.
These contrastive learning methods are typically employed as auxiliary tasks alongside the primary recommendation objective and have been demonstrated to enhance recommendation performance by improving user representation learning~\cite{qiu2022contrastive,yang2023debiased}.

Despite their empirical success, existing methods suffer from several limitations in contrastive pair construction, which may undermine their effectiveness in recommendation scenarios. 1) \textbf{Semantic Divergence}. Many existing methods construct contrastive pairs through random augmentation operations such as random masking \cite{xie2022contrastive,liu2021contrastive} and Dropout \cite{qiu2022contrastive}. However, in sequential recommendation where data is inherently sparse and exhibits sequential patterns, such random operations may lead to a complete change in the sequence’s semantics (i.e., user preferences). Bringing semantically different sequences closer together in embedding space may diminish the model's ability to discriminate among distinct user preferences.
Additionally, some methods determine contrastive pairs by clustering user representations derived from collaborative signals \cite{chen2022intent,qin2024intent}, where users within the same cluster are considered positive pairs. However, the sparse ID signals can lead to low-quality representations and inaccurate clustering results. 2) \textbf{Unlearnability}. Existing methods rely on predefined rules to construct positive pairs, such as directly selecting users from the same cluster \cite{chen2022intent,qin2024intent,liu2021contrastive}, or treating sequences sharing the same next item as positive pairs \cite{qiu2022contrastive,qin2024intent}.
These rigid heuristics impose strong assumptions that constrain models from autonomously learning optimal contrastive pairs. Moreover, the approach of using sequences with identical next items as positive pairs essentially replicates the recommendation objective (i.e., next-item prediction), providing no additional information gain.
Therefore, the suboptimal construction of contrastive pairs in existing methods limits their effectiveness and hinders contrastive learning's full potential.

Given these limitations, constructing high-quality contrastive samples remains a critical challenge. Semantic information, which is readily available in textual data such as product categories, brands, and descriptions, provides a promising solution. Unlike sparse behavior signals, semantic data maintains validity regardless of data volume or training dynamics, as it derives from structured knowledge rather than co-occurrence patterns \cite{zhou2020s3}. Additionally, semantic features offer complementary information beyond collaborative signals. Motivated by these advantages, we propose leveraging semantic information to construct superior contrastive pairs. However, accurately capturing user preferences requires models with powerful understanding and reasoning capabilities. Recent research has shown that large language models (LLMs) can effectively understand user preferences and achieve competitive performance on sequential recommendation tasks \cite{yang2024sequential}. Inspired by this, we propose to enhance contrastive learning through LLM-powered semantic retrieval.

In this paper, we propose SRA-CL (Semantic Retrieval-Augmented Contrastive Learning), a novel framework with two key innovations: 
1) \textbf{Semantic-based Retrieval}.
We develop a semantic-based retrieval mechanism that operates at both inter-user and intra-user levels.
For inter-user contrastive learning, we leverage LLMs to process sequential user interaction histories. Each sequence is fed to the LLM in chronological order of item interactions, where each item consists of both its attributes and textual description, enabling the model to generate preference-aware semantic embeddings through comprehensive understanding of user behavior patterns.
For intra-user contrastive learning, we enhance item understanding by providing LLMs with both item attributes and their contextual sequence information, producing context-aware semantic embeddings that capture both intrinsic item properties and their relevance within the recommendation context.
Subsequently, we leverage the semantic embeddings to retrieve the top-$k$ most similar users and items, constructing candidate positive sample pools for contrastive learning.
2) \textbf{Learnable Sample Synthesis}. To construct more effective contrastive samples, our framework incorporates a learnable sample synthesizer. For inter-user contrastive learning, the synthesizer dynamically generates positive samples for each user sequence by selectively combining elements from the candidate pool. This generation process is jointly optimized with the model training, ensuring the synthesized samples effectively improve representation learning.

Our main contributions are summarized as follows.
\begin{itemize}[leftmargin=1em]
\item We propose a model-agnostic framework, SRA-CL, which leverages semantic information and the capabilities of LLMs to construct better contrastive pairs, thereby improving the contrastive learning in sequential recommendation.

\item We propose a semantic-based retrieval approach for contrastive pair construction that integrates dual retrieval mechanisms: user retrieval for inter-user contrastive learning and item retrieval for intra-user contrastive learning, with each mechanism maintaining its dedicated candidate pool. To further enhance this framework, we introduce a learnable sample synthesizer that optimizes the contrastive sample generation process during model training.

\item We conduct extensive experiments on four public datasets to validate the superiority and model-agnostic nature of our approach, as well as to confirm the efficacy of each module.
\end{itemize}

\vspace{-0.5em}
\section{Preliminary}
\vspace{-0.5em}
 \subsection{Sequential Recommendation Task}
We denote the sets of users and items by $\mathcal{U}$ and $\mathcal{V}$, respectively. Each user $u\in \mathcal{U}$ has a chronological sequence of interacted items $\mathcal{S}_u = [v^u_1, v^u_2..., v^u_n]$, where $v^u_t$ indicates the item that $u$ interacted with at step $t$, and $n$ is the predefined maximum sequence length. For user sequences longer than $n$, we retain only the most recent $n$ items. The goal of sequential recommendation is to predict the next item $v^{+}$ according to $\mathcal{S}_u$, which can be formulated as:
 \begin{equation}
     \mathop{\arg\max}\limits_{v \in \mathcal{V}}  {P(v^{+}=v|\mathcal{S}_u)},
 \end{equation}
 where the probability $P$ represents the likelihood of item $v$ being the next item, conditioned on $\mathcal{S}_u$.
\vspace{-0.5em}
 \subsection{Sequential Recommendation Backbone} \label{backbone}
Our method is model-agnostic and can be integrated with various sequential recommendation models, as demonstrated in Section \ref{Validation of Model-Agnostic Characteristic}. To facilitate the introduction of our approach, we adopt the transformer architecture \cite{vaswani2017attention} as the backbone recommendation model following previous studies \cite{qin2023meta,qin2024intent,qiu2022contrastive}.
\begin{list}{}{
  \leftmargin=0em
  \itemindent=0em
  \topsep=0pt
  \parsep=3pt
  \itemsep=0pt
}
\item \textbf{Embedding Layer.} \label{emb layer}
We initialize an embedding matrix $\mathbf{M} \in \mathbb{R}^{|\mathcal{V}| \times d}$ to encode item IDs, where $|\mathcal{V}|$ represents the size of the item set and $d$ denotes the dimensionality of the latent space. Given a user interaction sequence $\mathcal{S}_u$, we obtain item embeddings $\mathbf{E}_u \in \mathbb{R}^{n \times d}$ and position embeddings $\mathbf{P} \in \mathbb{R}^{n \times d}$. Consequently, the input sequence $\mathcal{S}_u$ can be represented as $\mathbf{H}_u = \mathbf{E}_u + \mathbf{P}$.

\item \textbf{Sequence Encoder.}\label{user seq enc}
The representation of the input sequence is then fed into $L$ Transformer layers \cite{vaswani2017attention} to capture complex sequential patterns, which can be defined as follows:
\begin{equation} \label{rec_sequence_representation}
\mathbf{H}_u^{(L)} = \text{Transformer}(\mathbf{H}_u), \,\,\, \mathbf{h}_u = \mathbf{H}_u^{(L)}[-1].
\end{equation}
Here, $\mathbf{h}_u \in \mathbb{R}^{d}$ represents the last position of $\mathbf{H}_u^{(L)}$ and is selected as the final representation of $\mathcal{S}_u$.

\item \textbf{Prediction and Objective Function.}
During prediction, we calculate the probability of each item using \(\mathbf{\hat{y}} = \text{softmax}(\mathbf{h}_u \mathbf{M}^\mathrm{T})\), where \(\mathbf{\hat{y}} \in \mathbb{R}^{|\mathcal{V}|}\) and \(\hat{y}_v\) represents the likelihood of item \(v\) being the next item. 
For training, we adopt the same cross-entropy loss function as our baseline methods \cite{qin2023meta, qin2024intent, qiu2022contrastive} to ensure fairness, where $v^+$ denotes the ground truth item for user $u$.

\begin{equation} \label{rec_loss}
\mathcal{L}_{\text{Rec}} = -{\hat{y}}_{v^+} + \log\left(\sum_{v \in \mathcal{V}} \exp({\hat{y}}_v)\right).
\end{equation}

\end{list}



\vspace{-0.5em}
\section{The Framework of SRA-CL}
\begin{figure*}[t]
\setlength{\belowcaptionskip}{-2mm} 
  \centering
  \includegraphics[width=0.99\textwidth]{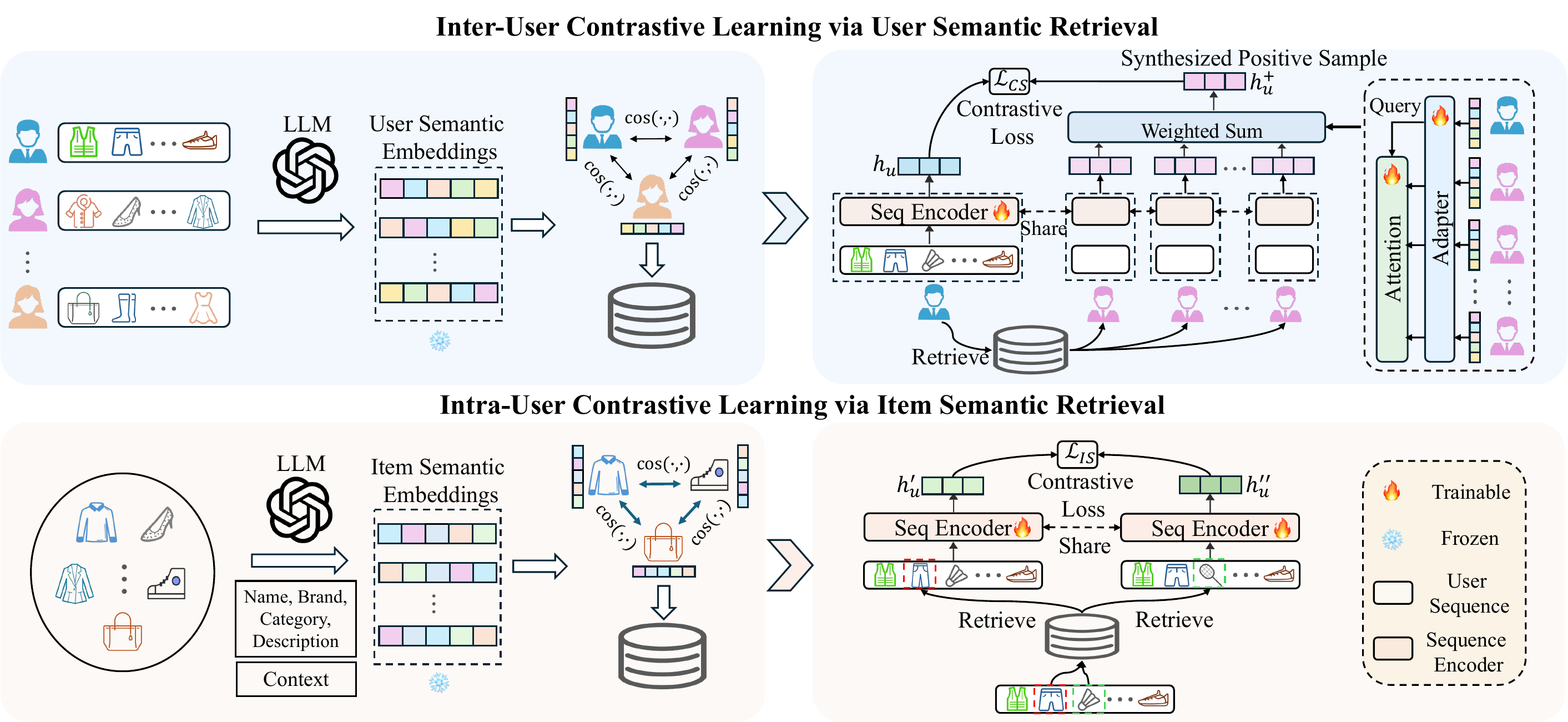}
  \caption{Overview of the proposed SRA-CL Framework.} 
   \label{method_fig}
\end{figure*}

In this section, we provide a detailed introduction to SRA-CL, which is shown in Figure \ref{method_fig}. SRA-CL integrates inter-user contrastive learning via user semantic retrieval and intra-user contrastive learning via item semantic retrieval. To further enhance the framework, we introduce a learnable sample synthesizer that optimizes the contrastive sample generation process during model training.

\vspace{-0.5em}
\subsection{Inter-User Contrastive Learning via User Semantic Retrieval}
\vspace{-0.5em}
SRA-CL employs semantic retrieval to generate reliable supervision signals for inter-user contrastive learning. Leveraging the advanced reasoning capabilities of LLMs, we first derive a comprehensive representation of user preferences, which are then encoded as semantic embeddings. Based on the similarity of these embeddings, we introduce a semantic-based retrieval mechanism to construct a candidate sample pool. Subsequently, a learnable contrastive sample synthesis method is employed to generate effective contrastive pairs.

\begin{list}{}{\leftmargin=0em}
\item \textbf{User Preference Understanding with LLMs.} \label{User Preference Understanding with LLMs}
Textual data (e.g., product categories, brands, and descriptions) plays a pivotal role in recommender systems by encoding rich semantic signals that reflect user preferences. Given user \( u \)'s interaction sequence \( \mathcal{S}_u \), we extract textual attributes for each item in chronological order, preserving both content and sequential context. These features are structured into a prompt \( \mathcal{P}_u \), where item attributes and their order explicitly guide the LLM in inferring user preferences \( \mathcal{A}_{u} = \text{LLM}(\mathcal{P}_u) \). The prompt template is detailed in Figure~\ref{prompt}.

Next, we employ a pretrained text embedding model $\mathcal{M}$ to extract and convert the semantic information contained in the textual responses of LLMs into embeddings, which is formatted as:
\begin{equation} \label{user_semantic_2}
    \tilde{\mathbf{h}}_u = \mathcal{M} (\mathcal{A}_u),
\end{equation}
where $\tilde{\mathbf{h}}_u \in \mathbb{R}^{\tilde{d}}$ represents the semantic embedding of user preferences and $\tilde{d}$ is the embedding size of the text embedding model $\mathcal{M}$. Specifically, $\mathcal{M}$ indicates SimCSE-RoBERTa \cite{gao2021simcse} in this paper due to its open-source availability and excellent sentence semantic extraction capabilities. The generated semantic embeddings are cached and remain fixed throughout the whole training process.

\begin{list}{}{\leftmargin=0em}
\item \textbf{Semantic-based User Retrieval.}
Once the semantic embeddings of user sequences are obtained, similar users can be retrieved based on semantic similarity. For a given user sequence \( \mathcal{S}_u \), we calculate the cosine similarity between its semantic embedding \( \tilde{\mathbf{h}}_u \) and the semantic embeddings of other users. Users are then ranked in descending order according to the computed semantic similarity. The top \( k \) users are retrieved to construct the homogeneous user pool for user \( u \), denoted as \( \mathcal{N}_u \).
\begin{equation} \label{retrieval_user}
\mathcal{N}_u = \{ u^\prime \in \mathcal{U} \setminus \{u\} \mid \text{rank}(\text{cosine\_similarity}(\tilde{\mathbf{h}}_u, \tilde{\mathbf{h}}_{u'})) \leq k \},
\end{equation}
where \( \mathcal{U} \setminus \{u\} \) denotes the set of all users except \( u \).
\item \textbf{Learnable Contrastive Sample Synthesis.}
Sole reliance on hard rules, such as selecting a user from the current user's dedicated candidate pool as the positive sample, often yields suboptimal solutions (as shown in Table \ref{ablation study}). To enhance contrastive sample construction, we introduce a learnable sample synthesizer that optimizes the contrastive sample generation process during model training.
Specifically, we first map the semantic representations of user sequences through a learnable adapter. Then, in the mapped space, we employ an attention mechanism, where the current user serves as a query to compute the probability \( p_{u,u'} \) that each candidate user \( u' \in \mathcal{N}_u \) is suitable as the positive sample for the current user \( u \). This process is formulated as:
\begin{align}
\label{generator_1}
    w_{u,u'} &= \text{LeakyReLU}(\mathbf{a}^\top [\mathbf{W}\tilde{\mathbf{h}}_u \Vert \mathbf{W}\tilde{\mathbf{h}}_{u'}]), \\
    \label{generator_2}
    p_{u,u'} &= \text{softmax}_{u'}(w_{u,u'}) = \frac{\exp(w_{u,u'})}{\sum_{u_k \in {\mathcal{N}_u}} \exp(w_{u, u_k})}, 
\end{align}

where $\mathbf{W} \in \mathbb{R}^{d \times \tilde{d}}$ is a learnable weight matrix, and $\Vert$ denotes the concatenation operation. $\mathbf{a} \in \mathbb{R}^{2d}$ represents a single-layer neural network used to generate the attention score, with the LeakyReLU activation function adopted \cite{velivckovic2017graph}. The softmax function is employed to transform the coefficients into probabilities.
Based on this, we generate the composite positive contrastive sample $\mathbf{h}_u^+$ for $\mathbf{h}_u$ by:

\begin{equation} \label{generator_3}
    \mathbf{h}_{u}^+ =  
    \sum\nolimits_{u'\in \mathcal{N}_u} p_{u,u'} \mathbf{h}_{u'}
   ,
\end{equation}
where $\mathbf{h}_{u'} \in \mathbb{R}^d$ is the recommendation model's output sequence representation for $u'$, as defined in Equation (\ref{rec_sequence_representation}).
This operation enables a fine-grained learnable selection of contrastive samples.
\end{list}

\item{\textbf{Inter-User Contrastive Loss.}}
For each user $u$, $\mathbf{h}_u$ is the sequence representation obtained from the recommendation model. The synthetic representation $\mathbf{h}_u^+$ is regarded as the positive sample for $\mathbf{h}_u$, while the remaining $N-1$ synthetic representations within the same batch are treated as negative samples for $\mathbf{h}_u$, where $N$ is the batch size.
We compute the inter-user contrastive loss $\mathcal{L}_\text{CS}$ as follows:
\begin{equation} \label{loss_cs}
\mathcal{L}_\text{CS} = -\log \frac{\exp\big({\mathbf{h}}_u\cdot {\mathbf{h}}_u^+)}
{{\exp\big({\mathbf{h}}_u\cdot {\mathbf{h}}_u^+)}+ \sum_{{\mathbf{h}}_u^- \in {\mathbf{H}}_u^-}\exp\big({\mathbf{h}}_u \cdot {\mathbf{h}}_u^-\big) },
\end{equation}
where $(\cdot)$ represents the inner product operation, $\mathbf{H}_u^-$ denotes the set of negative samples for $\mathbf{h}_u$. 

\end{list}

\subsection{Intra-User Contrastive Learning via Item Semantic Retrieval}

For intra-user contrastive learning, most existing methods apply predefined random perturbations to the original sequence to generate augmented views, which are treated as a pair of positive samples \cite{xie2022contrastive, qiu2022contrastive}. A significant limitation of them is the introduction of considerable uncertainty in the semantic similarity between positive samples. This substantial variation in user sequence semantics among positive samples undermines the reliability of the contrastive learning process.
To address this issue, we leverage a comprehensive understanding of both the semantic information of the item itself and the typical contexts in which the item appears. Based on this understanding, we replace certain items in the sequence with similar ones, resulting in semantic-consistent positive samples.

\begin{list}{}{\leftmargin=0em}
\vspace{-0.5em}
\item \textbf{Item Understanding with LLMs.} 
To enhance the LLM's comprehension of items, we provide two types of input information: (1) \textit{textual attributes of the item}, including category, brand, and description, which supply fundamental information and enable the LLM to perform a coarse-grained assessment of item similarity; and (2) \textit{user sequences containing the given item}. By analyzing the typical contexts in which an item appears, the LLM can infer the characteristics of its potential audience. This methodology facilitates more accurate evaluations of the relationships between items in the context of sequential recommendation. Given the token limit for LLM input, we have constrained the maximum number of item-related sequences in the prompt to 10, leaving the exploration of this value for future research.
Next, these two types of information for item $v$ are integrated into a structured prompt $\mathcal{P}_v$, which is processed by the language model to generate the item summary $\mathcal{A}_{v} = \text{LLM}(\mathcal{P}_v)$. The detailed prompt template is illustrated in Figure~\ref{prompt}.
Then, the pretrained text embedding model $\mathcal{M}$ is used to convert the textual responses of LLMs into embeddings: $\tilde{\mathbf{e}}_v = \mathcal{M} (\mathcal{A}_v)$.

\item{\textbf{Semantic-based Item Retrieval.}}
Similar to user retrieval, we compute the cosine similarity between the semantic embedding of an item and those of other items. Next, Top-$k$ most semantically relevant items for item $v$ are retrieved, which is formulated as:
\begin{equation} \label{retrival_item}
\mathcal{N}_v = \{ v' \in \mathcal{V} \setminus \{v\} \mid \text{rank}(\text{cosine\_similarity}(\tilde{\mathbf{e}}_v, \tilde{\mathbf{e}}_{v'})) \leq k \},
\end{equation}

\item{\textbf{Contrastive Sample Selection.}}
For intra-user contrastive learning, generating two semantic-consistent augmented views of the same user sequence is crucial. Here, we employ a semantic-based item substitution approach. Specifically, for each sequence $\mathcal{S}_u$, we randomly select 20\% of the items. For each selected item $v$, we substitute it with a semantically similar item sampled from its candidate pool $\mathcal{N}_v$. This operation yields an augmented sequence $\mathcal{S}^\prime_u$ derived from the original $\mathcal{S}_u$. By repeating this process, we obtain two augmented views, denoted as $\mathcal{S}^\prime_u$ and $\mathcal{S}^{\prime\prime}_u$, which form a positive sample pair.
Critically, the substitution is not entirely random but is guided by semantic similarity, which accounts for both item attribute similarity and contextual relevance in the recommendation scenario. This reduces uncertainty and enhances semantic consistency between augmented views.

Our preliminary experiments also explored the use of learnable synthesizers (analogous to inter-contrastive learning approaches) for generating substitute items, yet yielded no measurable performance improvements (shown in Table \ref{item learnable results}). 
This can be attributed to the inherently higher interpretability and quantifiability of item semantics relative to user preferences. Therefore, directly identifying appropriate substitutes from semantically similar candidate pools is simpler and more reliable compared to matching users with analogous preference patterns. A more detailed analysis is provided in Appendix \ref{Discussion on Learnable Sample Synthesis}.

\item{\textbf{Intra-User Contrastive Loss.}}
For the two augmented sequences $\mathcal{S}^\prime_u$ and $\mathcal{S}^{\prime\prime}_u$, we obtain their hidden vectors $\mathbf{h}^\prime_u$ and $\mathbf{h}^{\prime\prime}_u$ using the sequence encoder defined in Equation (\ref{rec_sequence_representation}). Then the intra-user contrastive loss can be calculated as:

\begin{equation} \label{loss_is}
\mathcal{L}_\text{IS} =-\log \frac{\exp\big(\mathbf{h}^\prime_u\cdot \mathbf{h}^{\prime\prime}_u)}
{\exp\big(\mathbf{h}^\prime_u\cdot \mathbf{h}^{\prime\prime}_u)+ \sum_{{\mathbf{h}}^{{\text{neg}}}_u \in {\mathbf{H}}^{\text{neg}}_u}\exp\big({\mathbf{h}}^\prime_u \cdot {\mathbf{h}}^{\text{neg}}_u\big) } 
\end{equation}

In a batch with a size of $N$, we have $2N$ augmented sequences. Among these, $\mathbf{h}^\prime_u$ and $\mathbf{h}^{\prime\prime}_u$ are positive samples of each other and are interchangeable. The remaining $2(N-1)$ samples excluding $\mathbf{h}^\prime_u$ and $\mathbf{h}^{\prime\prime}_u$ are considered negative samples ${\mathbf{H}}^{\text{neg}}_u$.

\end{list}

\begin{algorithm}
\caption{Training for SRA-CL}
\label{alg:train}
\begin{algorithmic}[1]
\REQUIRE Training data \(\{\mathcal{S}_u\}\) for all \(u \in \mathcal{U}\); hyperparameters \(\alpha\), \(\beta\), \(k\)
\STATE Obtain user semantic embeddings \(\{\tilde{\mathbf{h}}_u\}\) for all \(u \in \mathcal{U}\); obtain item semantic embeddings \(\{\tilde{\mathbf{e}}_v\}\) for all \(v \in \mathcal{V}\).
\STATE Freeze the embeddings \(\{\tilde{\mathbf{h}}_u\}\) and \(\{\tilde{\mathbf{e}}_v\}\), and initialize the model parameters.
\FOR{each iteration}
    \STATE Compute \(\mathbf{h}_u\) using Equation (\ref{rec_sequence_representation}).
    \STATE Calculate \(\mathbf{\hat{y}} = \text{softmax}(\mathbf{h}_u \mathbf{M}^\mathrm{T})\).
    \STATE Compute \(\mathcal{L}_{\text{Rec}}\) using Equation (\ref{rec_loss}).
    \STATE Retrieve \(\mathcal{N}_u\) for each \(u \in \mathcal{U}\) using Equation (\ref{retrieval_user}).
    \STATE Synthesize \(\mathbf{h}_u^+\) for each \(u\) using Equations (\ref{generator_1}), (\ref{generator_2}), and (\ref{generator_3}).
    \STATE Compute \(\mathcal{L}_{\text{CS}}\) using Equation (\ref{loss_cs}).
    \STATE Retrieve \(\mathcal{N}_v\) for each \(v \in \mathcal{V}\) using Equation (\ref{retrival_item}).
    \STATE Generate \(\mathcal{S}^\prime_u\) and \(\mathcal{S}^{\prime\prime}_u\), along with the corresponding \(\mathbf{h}^\prime_u\) and \(\mathbf{h}^{\prime\prime}_u\).
    \STATE Compute \(\mathcal{L}_{\text{IS}}\) using Equation (\ref{loss_is}).
    \STATE Calculate the total loss \(\mathcal{L} = \mathcal{L}_{\text{Rec}} + \alpha \mathcal{L}_{\text{CS}} + \beta \mathcal{L}_{\text{IS}}\).
    \STATE Update the model parameters using the gradient of \(\mathcal{L}\).
\ENDFOR
\STATE Return the final model parameters \(\theta\).
\end{algorithmic}
\end{algorithm}

\begin{algorithm}
\caption{Inference for SRA-CL}
\label{alg:inference}
\begin{algorithmic}[1]
\REQUIRE Trained model parameters \(\theta\); test data \(\{\mathcal{S}_u\}\)
\FOR{each user sequence in test data}
    \STATE Compute \(\mathbf{h}_u\) using Equation (\ref{rec_sequence_representation}).
    \STATE Calculate the predicted scores \(\mathbf{\hat{y}} = \text{softmax}(\mathbf{h}_u \mathbf{M}^\mathrm{T})\).
    \STATE Obtain the top-\(k\) items with the highest scores in \(\mathbf{\hat{y}}\).
\ENDFOR
\STATE Return the recommended items for all users.
\end{algorithmic}
\end{algorithm}

\subsection{Training and Inference}

During the training phase, all semantic embeddings are fixed. The training objective consists of three components: the loss of the recommendation model $\mathcal{L}_{\text{Rec}}$, which serves as the main loss, and the inter-user contrastive loss $\mathcal{L}_{\text{CS}}$ and intra-user contrastive loss $\mathcal{L}_{\text{IS}}$, which act as regularization terms.
\begin{equation}
\mathcal{L} = \mathcal{L}_{\text{Rec}} + \alpha \mathcal{L}_{\text{CS}} + \beta \mathcal{L}_{\text{IS}},
\end{equation}
where $\alpha$ and $\beta$ are hyperparameters.

During inference, only the recommendation backbone is utilized. The contrastive learning tasks and LLMs' semantic embeddings are not involved in the inference process. This implies that our framework can be deployed in real-world applications without incurring any additional inference latency from incorporating LLMs. 
The training and inference processes are detailed in Algorithm \ref{alg:train} and Algorithm \ref{alg:inference}, respectively.

\vspace{-0.6em}
\section{Experiments}
\vspace{-0.6em}
\subsection{Experimental Settings}
\begin{list}{}{\leftmargin=0em}
\item{\textbf{Datasets.}}
Following previous studies \cite{liu2021contrastive,xie2022contrastive,qin2023meta}, we conducted experiments on four public real-world datasets: Yelp, Amazon Sports, Beauty, and Office. The statistics for these datasets are presented in Table \ref{dataset}. More details about the datasets are shown in Appendix \ref{appendix:datasets}.


\item{\textbf{Evaluation Metrics.}}
To evaluate the performance of the models, we use widely recognized evaluation metrics: Hit Rate (HR) and Normalized Discounted Cumulative Gain (NDCG), follow previous studies ~\cite{zhang2024finerec,wang2024can,he2020lightgcn,kang2018self}.
The leave-one-out strategy is employed, where the last interaction is used for testing, the second-to-last interaction for validation, and the remaining interactions for training. To ensure an unbiased evaluation, we rank the prediction on the whole item set without sampling.

\item{\textbf{Baseline Methods.}}
We compare our method with {13} baseline methods, categorized into three groups: 1) \textit{classical methods} (GRU4Rec \cite{hidasi2015session}, SASRec \cite{kang2018self}, BERT4Rec \cite{sun2019bert4rec}), 2) \textit{contrastive learning-based methods} ($\rm {{S}^{3}\text{-}{Rec}}$ \cite{zhou2020s3}, CL4SRec \cite{xie2022contrastive}, CoSeRec \cite{liu2021contrastive}, ICLRec \cite{chen2022intent}, DuoRec \cite{qiu2022contrastive}, MCLRec \cite{qin2023meta}, ICSRec \cite{qin2024intent}), and 3) \textit{LLM-based methods} (LRD \cite{yang2024sequential}, RLMRec \cite{ren2024representation}, LLM-ESR \cite{liu2024llm}).

\item{\textbf{Implementation Details.}} All experiments are conducted with a single 32G V100 GPU. The embedding size is set to 64. 
We adopt the batch size of 256 and employ the Adam optimizer with a learning rate of 0.001. The dropout rate is set to 0.5 across all datasets. Following previous studies \cite{yang2024sequential}, we set the maximum sequence length to 20. The early stopping is applied if the metrics on the validation set do not improve over 10 consecutive epochs. 
For LLM, we use DeepSeek-V3 by invoking its API. We set the LLM's temperature $\tau$ to 0 and top-$p$ to 0.001. For the text embedding model $\mathcal{M}$, we use the pre-trained RoBERTa from Hugging Face. Note that identical settings are adopted for our method and baselines that involve LLMs and text embeddings to ensure fairness. More implementation details can be found in Appendix \ref{appendix:imple details}.
\end{list}


\begin{table*}[t]  
	\centering
\caption{Performance comparison of different methods on four datasets. Bold font indicates the best performance, while underlined values represent the second-best. SRA-CL achieves state-of-the-art results among all methods, as confirmed by a paired t-test with a significance level of 0.01. Due to space constraints, additional metrics (HR@10 and NDCG@10) are provided in Appendix \ref{Additional Comparison Results}.}
\label{comparison results}
 \resizebox{1.0\textwidth}{!}{
	\begin{tabular}{l|cc|cc|cc|cc}
		\toprule
		\multirow{2}*{Model} 
            & \multicolumn{2}{c|}{Yelp} 
            &\multicolumn{2}{c|}{Sports} 
            &\multicolumn{2}{c|}{Beauty} 
            &\multicolumn{2}{c}{Office}
            \\ 
    \cmidrule(lr){2-9} 
 
& HR@20 & NDCG@20 
& HR@20 & NDCG@20 
& HR@20 & NDCG@20  
& HR@20 & NDCG@20 
\\
\midrule 
GRU4Rec  &0.0639 &0.0243 &0.0325 &0.0129 &0.0488 &0.0189 &0.0956 &0.0361
\\
SASRec   &0.0899 &0.0390 &0.0498 &0.0216 &0.0887 &0.0382 &0.1329 &0.0482
\\
BERT4Rec  &0.0913 &0.0394 &0.0578 &0.0241 &0.0933 &0.0399 &0.1436 &0.0520
\\
\midrule
$\mathrm {\text{S}^3\text{-}\text{Rec}}$ &0.0964 &0.0443 &0.0607 &0.0262 &0.0994 &0.0414 &0.1568 &0.0571
\\
CL4SRec &0.0923 &0.0395 &0.0562 &0.0235 &0.0980 &0.0416 &0.1297 &0.0488
\\
CoSeRec  &0.0984 &0.0404 &0.0638 &0.0293 &0.1034 &0.0487 &0.1354 &0.0516
\\
ICLRec  &0.0974 &0.0432 &0.0636 &0.0284 &0.1056 &0.0482 &0.1513 &0.0559
\\
DuoRec  &\underline{0.1173} &0.0493 &0.0706 &0.0302 &0.1224 &0.0535 &0.1549 &0.0653 
\\
MCLRec &0.1150 &0.0486 &\underline{0.0736} &\underline{0.0318} &\underline{0.1239} &\underline{0.0536} &0.1629 &0.0684 
\\
ICSRec &0.1165 &\underline{0.0495} &0.0728 &0.0304 &0.1205 &0.0528 &\underline{0.1643} &\underline{0.0690}
\\
\midrule
LRD &0.1082 &0.0455 &0.0589 &0.0257 &0.0931 &0.0402 &0.1468 &0.0577
\\
RLMRec &0.1125 &0.0478 &0.0664 &0.0298 &0.1190 &0.0521 &0.1532 &0.0613
\\
LLM-ESR &0.1061 &0.0451 &0.0638 &0.0277 &0.1064 &0.0515 &0.1425 &0.0602
\\
\midrule
\textbf{SRA-CL} &\textbf{0.1282} &\textbf{0.0533} &\textbf{0.0823} &\textbf{0.0347} &\textbf{0.1314} &\textbf{0.0568} &\textbf{0.1702} &\textbf{0.0725}
\\
\midrule
Improvement &9.29\% &7.68\% &11.82\% &9.12\% &6.05\% &5.97\% &3.59\% &5.07\%
\\
\bottomrule
\end{tabular} }
\end{table*}

\subsection{Comparison Results with Baselines}

The comparison results are presented in Table \ref{comparison results}. Each experiment was conducted five times, and the average results are reported.
SRA-CL consistently outperforms all baseline methods across all datasets, achieving performance improvements of up to 11.82\%. The improvements are also confirmed by a paired t-test with a significance level of 0.01. 
Contrastive learning-based methods generally surpass traditional methods (GRU4Rec, SASRec, BERT4Rec). Among the contrastive learning baselines, MCLRec and ICSRec demonstrate superior performance. However, both methods underperform compared to SRA-CL, as they fail to control the quality of contrastive samples. SRA-CL mitigates this issue by introducing semantic-based retrieval augmentation, thereby improving the quality of contrastive samples and enhancing the overall effectiveness of contrastive learning.
Regarding LLM-enhanced baselines, they demonstrate superior results compared to classical methods. However, our proposed SRA-CL achieves significant improvements over these LLM approaches. Unlike existing LLM-based methods, SRA-CL is fundamentally different in motivation---it specifically addresses the limitations in contrastive learning through enhanced construction of positive sample pairs using semantic information.

\subsection{Validation of Model-Agnostic Characteristic}
\label{Validation of Model-Agnostic Characteristic}
\begin{figure*}[h]
\setlength{\abovecaptionskip}{-0.1mm} 
\setlength{\belowcaptionskip}{-5mm} 
  \centering
  \includegraphics[width=0.95\textwidth]{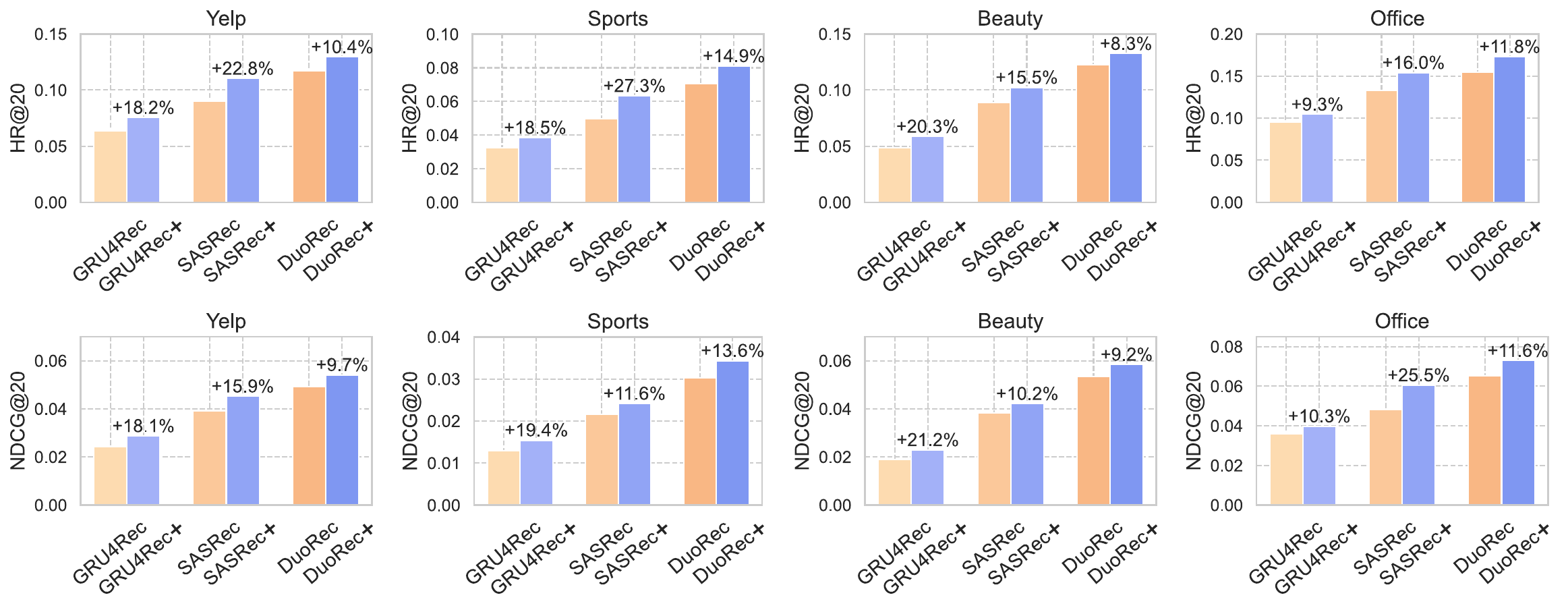}
  \caption{Experimental results demonstrating the model-agnostic nature and strong generalization capability of SRA-CL. ``+'' indicates the addition of SRA-CL to different recommendation models.
  } 
   \label{diejia_experiments}
\end{figure*}
In this section, we validate the model-agnostic nature of our method. We select three classic recommendation models (GRU4Rec, SASRec, DuoRec) as the backbone and integrate SRA-CL to observe performance changes. 
We retain the original loss functions of the backbones and introduce our contrastive loss $\mathcal{L}_{CS}$ and $\mathcal{L}_{IS}$ during training.
The results are shown in Figure \ref{diejia_experiments}, which indicate that for all three backbone methods, the versions enhanced with SRA-CL ("+") consistently outperform the original versions. Specifically, HR@20 improves by 8.3\% to 27.3\%, and NDCG@20 increases by 9.7\% to 25.5\%. These findings validate that SRA-CL can robustly improve the performance of various recommendation models.

\vspace{-0.6em}
\subsection{Ablation Study}
\vspace{-0.5em}

\begin{table}
\renewcommand\arraystretch{0.9}
\centering
\caption{Ablation study on all datasets.}
\label{ablation study}
\setlength\tabcolsep{3.8pt}
\resizebox{0.75\textwidth}{!}{
\begin{tabular}{l|c|ccccccc}
\toprule
 & Metric & {\makecell[c]{w/o \\CL}}& {\makecell[c]{w/o \\ $\mathcal{L}_{CS}$}} & {\makecell[c]{w/o \\$\mathcal{L}_{IS}$}} & {\makecell[c]{w/o \\ learn.}} & {\makecell[c]{w/o \\ semantic}} & {\makecell[c]{w/o \\LLM}} & \textbf{Ours}  \\
\midrule
\multirow{2}*{Yelp}  
&H@20 &0.1101 & 0.1203 & 0.1228 & 0.1253 &0.1187 &0.1190 & \textbf{0.1282}  \\
&N@20 &0.0473 &0.0504  &0.0519  &0.0520  &0.0495 &0.0501 &\textbf{0.0533}   \\
\midrule
\multirow{2}*{Sports}  
&H@20 &0.0745 & 0.0780 & 0.0795 & 0.0792 &0.0772 &0.0781 & \textbf{0.0823}  \\
&N@20 &0.0296 &0.0315  &0.0332  &0.0336  &0.0311 &0.0314 &\textbf{0.0347}   \\
\midrule
\multirow{2}*{Beauty}  
&H@20 &0.1206 & 0.1273 & 0.1279 & 0.1273 &0.1265 &0.1259 & \textbf{0.1314}  \\
&N@20 &0.0518 &0.0546  &0.0545  &0.0551  &0.0532 &0.0537 &\textbf{0.0568}  \\

\midrule
\multirow{2}*{Office}  
&H@20 &0.1476 & 0.1621 & 0.1619 & 0.1617 &0.1624 &0.1643 & \textbf{0.1702}  \\
&N@20 &0.0599 &0.0691  &0.0689  &0.0681  &0.0673 &0.0692 &\textbf{0.0725}   \\
\bottomrule
\end{tabular}
}
\end{table}

In this section, we evaluate the effectiveness of each component in SRA-CL. The results, presented in Table~\ref{ablation study}, demonstrate the impact of removing individual modules.
Overall, the results show that removing any component degrades model performance, confirming the necessity of each module. Specifically, the variants ``w/o $\mathcal{L}_{CS}$'' and ``w/o $\mathcal{L}_{IS}$'' exhibit significant performance drops, highlighting the importance of both inter-user and intra-user contrastive learning objectives. The ``w/o CL'' variant suffers a more severe performance decline than those removing only one contrastive objective, suggesting that these two types of objectives complement each other.
Additionally, the ``w/o learn.'' variant also leads to reduced performance, indicating that a learning-based sample synthesizer is more effective than random selection for inter-user contrastive learning.
Furthermore, removing semantic information and relying solely on collaborative signals for retrieval (``w/o semantic'') results in a notable performance decline, underscoring the importance of semantic information in constructing high-quality contrastive samples. This finding aligns with our initial motivation. Similarly, the absence of LLM-based text processing (``w/o LLM'') also results in performance degradation, demonstrating that utilizing the LLM's ability to understand and reason about user preferences is crucial.


\begin{figure}[t]
\setlength{\abovecaptionskip}{-0.1mm} 
\setlength{\belowcaptionskip}{-5mm} 
  \centering
  \includegraphics[width=0.85\columnwidth]{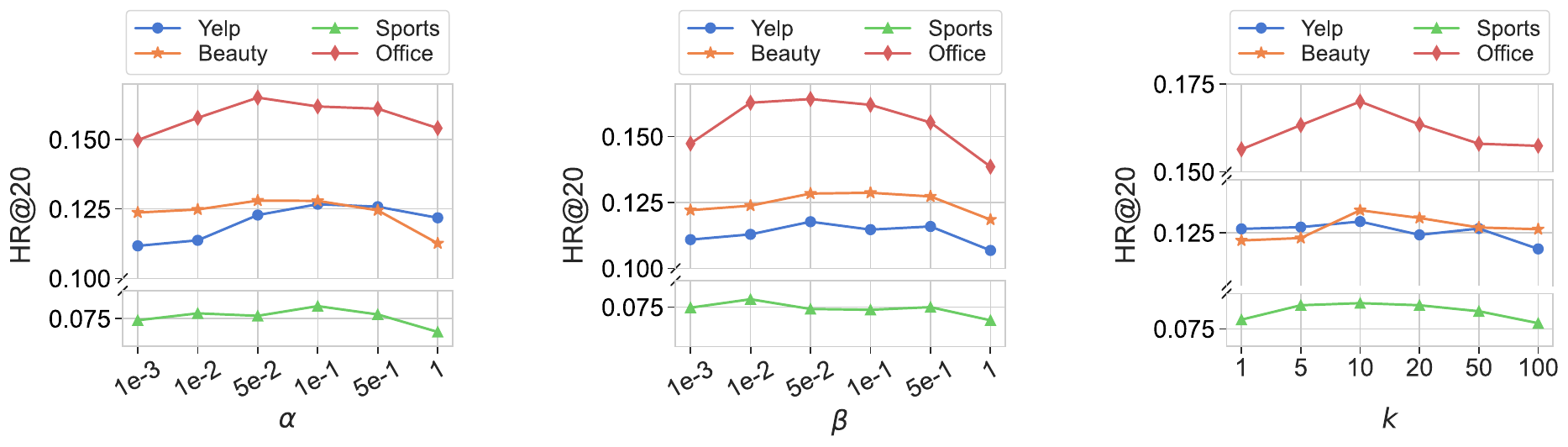}
  \caption{Hyperparameter experiments on the weight of $\mathcal{L}_{CS}$ ($\alpha$), the weight of $\mathcal{L}_{IS}$ ($\beta$), and the number of retrieved users/items ($k$).} 
   \label{param_study_loss}
\end{figure}

\vspace{-0.5em}
\subsection{Hyperparameter Study}
\vspace{-0.5em}
In this section, we investigate the impacts of three key hyperparameters, $\alpha$, $\beta$, and $k$. Here, $\alpha$ and $\beta$ are the weights of $\mathcal{L}_{CS}$ and $\mathcal{L}_{IS}$, respectively, while $k$ denotes the number of retrieved users/items.
From Figure \ref{param_study_loss}, we observe that as both $\alpha$ and $\beta$ increase, the model's performance initially improves slightly and then decreases marginally. 
Empirically, the optimal range for $\alpha$ and $\beta$ is between 0.05 and 0.1. This is reasonable as contrastive learning loss acts as a regularization term.
As the value of $k$ increases, the performance initially improves and then declines, with the optimal value around 10. 
As $k$ increases, the semantic relevance of retrieved neighbors decreases and randomness increases. A very small $k$ results in a candidate set that is too small without diversity. Conversely, a very large $k$ loses semantic relevance, thereby degrading the effectiveness of contrastive learning. Note that NDCG@20 results are provided in Figure \ref{param_study_loss_ndcg} due to space limitation.

\vspace{-0.5em}
\subsection{Contrastive Learning in Sparse Data: Analyzing SRA-CL’s Superiority}
\vspace{-0.5em}
\begin{wrapfigure}{r}{0.5\textwidth}
\setlength{\abovecaptionskip}{-0.1mm} 
\setlength{\belowcaptionskip}{-2mm} 
  \centering
  \includegraphics[width=0.5\textwidth]{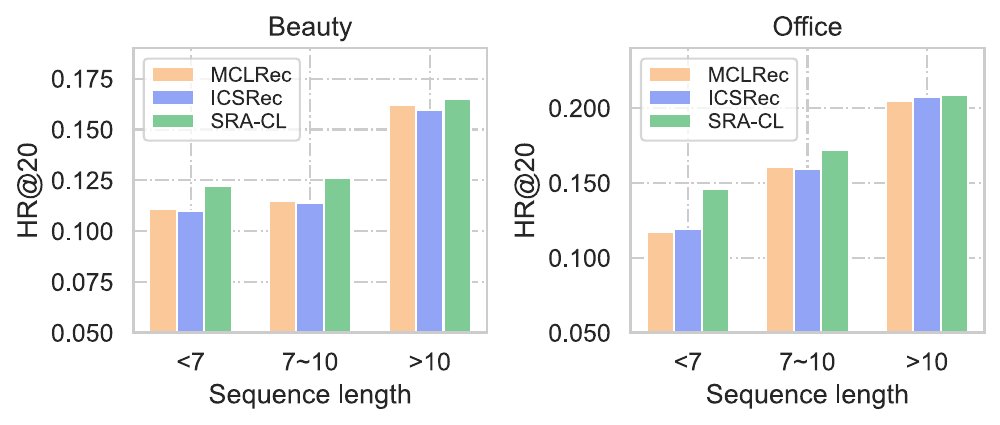}
  \caption{Performance comparison on different user groups
among MCLRec, ICSRec and Ours.} 
   \label{group_study}
\end{wrapfigure}
To further examine SRA-CL’s capability in mitigating the issue of low-quality contrastive samples in data-sparse scenarios, we categorize 
user sequences into three groups based on their length and compare the evaluation results of different methods. Due to space limitation, we present the experimental results for Beauty and Office, as shown in Figure \ref{group_study}. 
By comparing SRA-CL with the two strongest contrastive learning baselines (MCLRec and ICSRec), we observe that SRA-CL consistently outperforms them across all user groups. 
Notably, our method achieves greater improvements in sparser user groups (e.g., those with fewer than 7 or 7-10 interactions). This result further validates our motivation: while MCLRec and ICSRec construct contrastive sample pairs based on collaborative signals, their performance degrades in data-sparse scenarios due to the diminished quality of contrastive samples. In contrast, our method significantly enhances the quality of contrastive pairs by incorporating semantic information, leading to superior performance under sparse data conditions.

\vspace{-0.7em}
\section{Related Work}
\vspace{-0.7em}
\begin{list}{}{\leftmargin=0em}
\item \textbf{{Contrastive Learning in Sequential Recommendation.}}
Contrastive learning has been successfully used to enhance sequential recommendation~\cite{zhang2024recdcl, jiang2023diffkg, yu2022graph, wu2021self, xia2023automated, yin2022autogcl,cui2022dual,cui2024context,qin2024intent}.
In terms of the composition of contrastive samples, we categorize existing methods into two types:
(1) Inter-user. This involves generating contrastive samples from different user sequences. For example, ICLRec~\cite{chen2022intent} clusters user interests into distinct categories by K-Means and brings the representations of users with similar interests closer. ICSRec \cite{qin2024intent} further segments a user's behavior sequence into multiple subsequences to generate finer-grained user intentions for contrastive learning. These methods generate contrastive supervision signals based on collaborative signals. However, the sparsity of the co-occurrence pattern leads to unreliable clustering results, which in turn affects the performance of contrastive learning.
(2) Intra-user. This involves applying perturbations to
the original sequence to generate augmented views. The two views of the same sequence are treated as a pair of positive samples. For example, CL4SRec~\cite{xie2022contrastive} employs three data-level augmentation operators: Cropping, Masking, and Reordering, to create contrastive pairs. 
CoSeRec~\cite{liu2021contrastive} introduces two additional informative augmentation operators, building upon the foundation of CL4SRec. 
In addition, some methods generate augmented views from the model's hidden layers. A notable example is DuoRec~\cite{qiu2022contrastive}, which creates positive pairs by forward-passing a sequence representation twice with different dropout masks. MCLRec~\cite{qin2023meta} further combines data-level and model-level augmentation. Despite their effectiveness, they employ random operators, introducing significant uncertainty and potentially generating unreasonable positive samples for contrastive learning.

\item \textbf{Sequential Recommendation with LLMs.}
Building upon foundations laid by traditional recommender systems \cite{he2020lightgcn,li2023alpt,li2024optfp}, recent studies have successfully integrated LLMs into the recommender paradigm~\cite{wu2024survey,zhao2023recommender,li2024large,cui2025comprehending}. 
Overall, LLMs are employed either as direct recommenders or as tools for extracting semantic information~\cite{li2023prompt,liao2024llara,zhao2024let,boz2024improving,ren2024enhancing}. In the former approach, all inputs are converted into textual format, and the LLM generates recommendations based on its pre-trained knowledge or after undergoing supervised fine-tuning. Representative examples include LC-Rec~\cite{zheng2024adapting}, LLM-TRSR~\cite{zheng2024harnessing}, and CALRec~\cite{li2024calrec}. However, these methods rely on the inference process of large language models to generate recommendation results, which is computationally expensive and often challenging to deploy in practical scenarios. Another line of research~\cite{ren2024enhancing,wang2024can,liu2024llm,yang2024sequential,zhao2024let,boz2024improving} leverages LLMs to process semantic information and incorporates it into traditional ID-based models. 
For example, SLIM~\cite{wang2024can} distills knowledge from large-scale LLMs into a smaller student LLM to improve the recommendation model. LLM-ESR~\cite{liu2024llm} addresses the long-tail problem by leveraging collaborative signals and semantic information through dual-view modeling and self-distillation. LRD~\cite{yang2024sequential} utilizes the LLM to explore potential relations between items and reconstructs one item based on its relation to another.
Unlike the aforementioned methods, our approach, grounded in the essence of contrastive learning, aims to construct more effective contrastive pairs with LLMs.
\end{list}

\vspace{-0.8em}
\section{Conclusion}
\vspace{-0.7em}
In this paper, we analyze the limitations of contrastive learning in sequential recommendation, namely Semantic Divergence and Unlearnability. To address these issues, we propose SRA-CL, a novel framework that enhances contrastive sample construction by integrating LLM-based semantic retrieval with a learnable sample synthesizer. SRA-CL leverages the capabilities of LLMs without increasing the inference time of the recommendation model, making it practical for large-scale real-world applications. Through comprehensive experiments, we demonstrate that LLM-based semantic-guided contrastive sample construction improves the contrastive learning, and we validate the effectiveness of the learnable sample synthesis mechanism. Furthermore, experiments with different recommendation model backbones confirm the model-agnostic nature of our approach. 

\vspace{-0.5em}
\section{Acknowledgements}
\vspace{-0.7em}
This work was supported by the Early Career Scheme (No. CityU 21219323) and the General Research Fund (No. CityU 11220324) of the University Grants Committee (UGC), the NSFC Young Scientists Fund (No. 9240127), and the Donation for Research Projects (No. 9220187 and No. 9229164).

\bibliographystyle{plainnat} 
\bibliography{reference}

@inproceedings{qin2023meta,
author = {Qin, Xiuyuan and Yuan, Huanhuan and Zhao, Pengpeng and Fang, Junhua and Zhuang, Fuzhen and Liu, Guanfeng and Liu, Yanchi and Sheng, Victor},
title = {Meta-optimized Contrastive Learning for Sequential Recommendation},
year = {2023},
booktitle = {Proceedings of the 46th International ACM SIGIR Conference on Research and Development in Information Retrieval},
pages = {89–98},
}

@inproceedings{qiu2022contrastive,
  title={Contrastive learning for representation degeneration problem in sequential recommendation},
  author={Qiu, Ruihong and Huang, Zi and Yin, Hongzhi and Wang, Zijian},
  booktitle={Proceedings of the fifteenth ACM international conference on web search and data mining},
  pages={813--823},
  year={2022}
}

@inproceedings{xie2022contrastive,
  title={Contrastive learning for sequential recommendation},
  author={Xie, Xu and Sun, Fei and Liu, Zhaoyang and Wu, Shiwen and Gao, Jinyang and Zhang, Jiandong and Ding, Bolin and Cui, Bin},
  booktitle={2022 IEEE 38th international conference on data engineering (ICDE)},
  pages={1259--1273},
  year={2022},
  organization={IEEE}
}

@inproceedings{kang2018self,
  title={Self-attentive sequential recommendation},
  author={Kang, Wang-Cheng and McAuley, Julian},
  booktitle={2018 IEEE international conference on data mining (ICDM)},
  pages={197--206},
  year={2018},
  organization={IEEE}
}

@inproceedings{chen2022intent,
  title={Intent contrastive learning for sequential recommendation},
  author={Chen, Yongjun and Liu, Zhiwei and Li, Jia and McAuley, Julian and Xiong, Caiming},
  booktitle={Proceedings of the ACM Web Conference 2022},
  pages={2172--2182},
  year={2022}
}

@inproceedings{zhou2023equivariant,
  title={Equivariant contrastive learning for sequential recommendation},
  author={Zhou, Peilin and Gao, Jingqi and Xie, Yueqi and Ye, Qichen and Hua, Yining and Kim, Jaeboum and Wang, Shoujin and Kim, Sunghun},
  booktitle={Proceedings of the 17th ACM Conference on Recommender Systems},
  pages={129--140},
  year={2023}
}

@inproceedings{yang2023debiased,
  title={Debiased Contrastive Learning for Sequential Recommendation},
  author={Yang, Yuhao and Huang, Chao and Xia, Lianghao and Huang, Chunzhen and Luo, Da and Lin, Kangyi},
  booktitle={Proceedings of the ACM Web Conference 2023},
  pages={1063--1073},
  year={2023}
}

@article{hidasi2015session,
  title={Session-based recommendations with recurrent neural networks},
  author={Hidasi, Bal{\'a}zs and Karatzoglou, Alexandros and Baltrunas, Linas and Tikk, Domonkos},
  journal={arXiv preprint arXiv:1511.06939},
  year={2015}
}

@inproceedings{zhou2020s3,
  title={S3-rec: Self-supervised learning for sequential recommendation with mutual information maximization},
  author={Zhou, Kun and Wang, Hui and Zhao, Wayne Xin and Zhu, Yutao and Wang, Sirui and Zhang, Fuzheng and Wang, Zhongyuan and Wen, Ji-Rong},
  booktitle={Proceedings of the 29th ACM international conference on information \& knowledge management},
  pages={1893--1902},
  year={2020}
}

@inproceedings{sun2019bert4rec,
  title={BERT4Rec: Sequential recommendation with bidirectional encoder representations from transformer},
  author={Sun, Fei and Liu, Jun and Wu, Jian and Pei, Changhua and Lin, Xiao and Ou, Wenwu and Jiang, Peng},
  booktitle={Proceedings of the 28th ACM international conference on information and knowledge management},
  pages={1441--1450},
  year={2019}
}

@article{liu2021contrastive,
  title={Contrastive self-supervised sequential recommendation with robust augmentation},
  author={Liu, Zhiwei and Chen, Yongjun and Li, Jia and Yu, Philip S and McAuley, Julian and Xiong, Caiming},
  journal={arXiv preprint arXiv:2108.06479},
  year={2021}
}

@article{devlin2018bert,
  title={Bert: Pre-training of deep bidirectional transformers for language understanding},
  author={Devlin, Jacob and Chang, Ming-Wei and Lee, Kenton and Toutanova, Kristina},
  journal={arXiv preprint arXiv:1810.04805},
  year={2018}
}

@article{vaswani2017attention,
  title={Attention is all you need},
  author={Vaswani, Ashish and Shazeer, Noam and Parmar, Niki and Uszkoreit, Jakob and Jones, Llion and Gomez, Aidan N and Kaiser, {\L}ukasz and Polosukhin, Illia},
  journal={Advances in neural information processing systems},
  volume={30},
  year={2017}
}

@inproceedings{chen2020simple,
  title={A simple framework for contrastive learning of visual representations},
  author={Chen, Ting and Kornblith, Simon and Norouzi, Mohammad and Hinton, Geoffrey},
  booktitle={International conference on machine learning},
  pages={1597--1607},
  year={2020},
  organization={PMLR}
}

@article{jiang2023diffkg,
  title={DiffKG: Knowledge Graph Diffusion Model for Recommendation},
  author={Jiang, Yangqin and Yang, Yuhao and Xia, Lianghao and Huang, Chao},
  journal={arXiv preprint arXiv:2312.16890},
  year={2023}
}

@inproceedings{yu2022graph,
  title={Are graph augmentations necessary? simple graph contrastive learning for recommendation},
  author={Yu, Junliang and Yin, Hongzhi and Xia, Xin and Chen, Tong and Cui, Lizhen and Nguyen, Quoc Viet Hung},
  booktitle={Proceedings of the 45th international ACM SIGIR conference on research and development in information retrieval},
  pages={1294--1303},
  year={2022}
}

@inproceedings{wu2021self,
  title={Self-supervised graph learning for recommendation},
  author={Wu, Jiancan and Wang, Xiang and Feng, Fuli and He, Xiangnan and Chen, Liang and Lian, Jianxun and Xie, Xing},
  booktitle={Proceedings of the 44th international ACM SIGIR conference on research and development in information retrieval},
  pages={726--735},
  year={2021}
}

@inproceedings{xia2023automated,
  title={Automated Self-Supervised Learning for Recommendation},
  author={Xia, Lianghao and Huang, Chao and Huang, Chunzhen and Lin, Kangyi and Yu, Tao and Kao, Ben},
  booktitle={Proceedings of the ACM Web Conference 2023},
  pages={992--1002},
  year={2023}
}

@inproceedings{yin2022autogcl,
  title={Autogcl: Automated graph contrastive learning via learnable view generators},
  author={Yin, Yihang and Wang, Qingzhong and Huang, Siyu and Xiong, Haoyi and Zhang, Xiang},
  booktitle={Proceedings of the AAAI conference on artificial intelligence},
  volume={36},
  number={8},
  pages={8892--8900},
  year={2022}
}

@inproceedings{he2020lightgcn,
  title={Lightgcn: Simplifying and powering graph convolution network for recommendation},
  author={He, Xiangnan and Deng, Kuan and Wang, Xiang and Li, Yan and Zhang, Yongdong and Wang, Meng},
  booktitle={Proceedings of the 43rd International ACM SIGIR conference on research and development in Information Retrieval},
  pages={639--648},
  year={2020}
}

@inproceedings{cui2022dual,
  title={Dual Disentangled Attention for Multi-Information Utilization in Sequential Recommendation},
  author={Cui, Ziqiang and Su, Yixin and Lin, Fangquan and Yang, Cheng and Zhang, Hanwei and Zhang, Jihai},
  booktitle={2022 International Joint Conference on Neural Networks (IJCNN)},
  pages={1--8},
  year={2022},
  organization={IEEE}
}

@inproceedings{zhang2024recdcl,
  title={RecDCL: Dual Contrastive Learning for Recommendation},
  author={Zhang, Dan and Geng, Yangliao and Gong, Wenwen and Qi, Zhongang and Chen, Zhiyu and Tang, Xing and Shan, Ying and Dong, Yuxiao and Tang, Jie},
  booktitle={Proceedings of the ACM on Web Conference 2024},
  pages={3655--3666},
  year={2024}
}

@article{wu2024survey,
  title={A survey on large language models for recommendation},
  author={Wu, Likang and Zheng, Zhi and Qiu, Zhaopeng and Wang, Hao and Gu, Hongchao and Shen, Tingjia and Qin, Chuan and Zhu, Chen and Zhu, Hengshu and Liu, Qi and others},
  journal={World Wide Web},
  volume={27},
  number={5},
  pages={60},
  year={2024},
  publisher={Springer}
}

@article{zhao2023recommender,
  title={Recommender systems in the era of large language models (llms)},
  author={Zhao, Zihuai and Fan, Wenqi and Li, Jiatong and Liu, Yunqing and Mei, Xiaowei and Wang, Yiqi and Wen, Zhen and Wang, Fei and Zhao, Xiangyu and Tang, Jiliang and others},
  journal={arXiv preprint arXiv:2307.02046},
  year={2023}
}

@inproceedings{li2024large,
  title={Large Language Models for Generative Recommendation: A Survey and Visionary Discussions},
  author={Li, Lei and Zhang, Yongfeng and Liu, Dugang and Chen, Li},
  booktitle={Proceedings of the 2024 Joint International Conference on Computational Linguistics, Language Resources and Evaluation (LREC-COLING 2024)},
  pages={10146--10159},
  year={2024}
}

@inproceedings{zheng2024harnessing,
  title={Harnessing large language models for text-rich sequential recommendation},
  author={Zheng, Zhi and Chao, Wenshuo and Qiu, Zhaopeng and Zhu, Hengshu and Xiong, Hui},
  booktitle={Proceedings of the ACM on Web Conference 2024},
  pages={3207--3216},
  year={2024}
}

@inproceedings{li2024calrec,
  title={Calrec: Contrastive alignment of generative llms for sequential recommendation},
  author={Li, Yaoyiran and Zhai, Xiang and Alzantot, Moustafa and Yu, Keyi and Vuli{\'c}, Ivan and Korhonen, Anna and Hammad, Mohamed},
  booktitle={Proceedings of the 18th ACM Conference on Recommender Systems},
  pages={422--432},
  year={2024}
}

@inproceedings{li2023prompt,
  title={Prompt distillation for efficient llm-based recommendation},
  author={Li, Lei and Zhang, Yongfeng and Chen, Li},
  booktitle={Proceedings of the 32nd ACM International Conference on Information and Knowledge Management},
  pages={1348--1357},
  year={2023}
}

@inproceedings{liao2024llara,
  title={Llara: Large language-recommendation assistant},
  author={Liao, Jiayi and Li, Sihang and Yang, Zhengyi and Wu, Jiancan and Yuan, Yancheng and Wang, Xiang and He, Xiangnan},
  booktitle={Proceedings of the 47th International ACM SIGIR Conference on Research and Development in Information Retrieval},
  pages={1785--1795},
  year={2024}
}

@inproceedings{zhao2024let,
  title={Let me do it for you: Towards llm empowered recommendation via tool learning},
  author={Zhao, Yuyue and Wu, Jiancan and Wang, Xiang and Tang, Wei and Wang, Dingxian and de Rijke, Maarten},
  booktitle={Proceedings of the 47th International ACM SIGIR Conference on Research and Development in Information Retrieval},
  pages={1796--1806},
  year={2024}
}

@article{boz2024improving,
  title={Improving sequential recommendations with llms},
  author={Boz, Artun and Zorgdrager, Wouter and Kotti, Zoe and Harte, Jesse and Louridas, Panos and Jannach, Dietmar and Fragkoulis, Marios},
  journal={arXiv preprint arXiv:2402.01339},
  year={2024}
}

@inproceedings{zheng2024adapting,
  title={Adapting large language models by integrating collaborative semantics for recommendation},
  author={Zheng, Bowen and Hou, Yupeng and Lu, Hongyu and Chen, Yu and Zhao, Wayne Xin and Chen, Ming and Wen, Ji-Rong},
  booktitle={2024 IEEE 40th International Conference on Data Engineering (ICDE)},
  pages={1435--1448},
  year={2024},
  organization={IEEE}
}

@inproceedings{wang2024can,
  title={Can Small Language Models be Good Reasoners for Sequential Recommendation?},
  author={Wang, Yuling and Tian, Changxin and Hu, Binbin and Yu, Yanhua and Liu, Ziqi and Zhang, Zhiqiang and Zhou, Jun and Pang, Liang and Wang, Xiao},
  booktitle={Proceedings of the ACM on Web Conference 2024},
  pages={3876--3887},
  year={2024}
}

@inproceedings{ren2024enhancing,
  title={Enhancing sequential recommenders with augmented knowledge from aligned large language models},
  author={Ren, Yankun and Chen, Zhongde and Yang, Xinxing and Li, Longfei and Jiang, Cong and Cheng, Lei and Zhang, Bo and Mo, Linjian and Zhou, Jun},
  booktitle={Proceedings of the 47th International ACM SIGIR Conference on Research and Development in Information Retrieval},
  pages={345--354},
  year={2024}
}

@inproceedings{liu2024llm,
  title={LLM-ESR: Large Language Models Enhancement for Long-tailed Sequential Recommendation},
  author={Liu, Qidong and Wu, Xian and Wang, Yejing and Zhang, Zijian and Tian, Feng and Zheng, Yefeng and Zhao, Xiangyu},
  booktitle={The Thirty-eighth Annual Conference on Neural Information Processing Systems},
  year={2024}
}

@inproceedings{yang2024sequential,
  title={Sequential recommendation with latent relations based on large language model},
  author={Yang, Shenghao and Ma, Weizhi and Sun, Peijie and Ai, Qingyao and Liu, Yiqun and Cai, Mingchen and Zhang, Min},
  booktitle={Proceedings of the 47th International ACM SIGIR Conference on Research and Development in Information Retrieval},
  pages={335--344},
  year={2024}
}

@inproceedings{qin2024intent,
  title={Intent Contrastive Learning with Cross Subsequences for Sequential Recommendation},
  author={Qin, Xiuyuan and Yuan, Huanhuan and Zhao, Pengpeng and Liu, Guanfeng and Zhuang, Fuzhen and Sheng, Victor S},
  booktitle={Proceedings of the 17th ACM International Conference on Web Search and Data Mining},
  pages={548--556},
  year={2024}
}

@inproceedings{zhang2024finerec,
  title={FineRec: Exploring Fine-grained Sequential Recommendation},
  author={Zhang, Xiaokun and Xu, Bo and Wu, Youlin and Zhong, Yuan and Lin, Hongfei and Ma, Fenglong},
  booktitle={Proceedings of the 47th International ACM SIGIR Conference on Research and Development in Information Retrieval},
  pages={1599--1608},
  year={2024}
}

@inproceedings{gao2021simcse,
  title={{SimCSE}: Simple Contrastive Learning of Sentence Embeddings},
  author={Gao, Tianyu and Yao, Xingcheng and Chen, Danqi},
  booktitle={Empirical Methods in Natural Language Processing (EMNLP)},
  year={2021}
}

@inproceedings{cui2024context,
  title={Context Matters: Enhancing Sequential Recommendation with Context-aware Diffusion-based Contrastive Learning},
  author={Cui, Ziqiang and Wu, Haolun and He, Bowei and Cheng, Ji and Ma, Chen},
  booktitle={Proceedings of the 33rd ACM International Conference on Information and Knowledge Management},
  pages={404--414},
  year={2024}
}

@article{velivckovic2017graph,
  title={Graph attention networks},
  author={Veli{\v{c}}kovi{\'c}, Petar and Cucurull, Guillem and Casanova, Arantxa and Romero, Adriana and Lio, Pietro and Bengio, Yoshua},
  journal={arXiv preprint arXiv:1710.10903},
  year={2017}
}

@inproceedings{li2023multi,
  title={Multi-intention oriented contrastive learning for sequential recommendation},
  author={Li, Xuewei and Sun, Aitong and Zhao, Mankun and Yu, Jian and Zhu, Kun and Jin, Di and Yu, Mei and Yu, Ruiguo},
  booktitle={Proceedings of the sixteenth ACM international conference on web search and data mining},
  pages={411--419},
  year={2023}
}

@article{yu2023self,
  title={Self-supervised learning for recommender systems: A survey},
  author={Yu, Junliang and Yin, Hongzhi and Xia, Xin and Chen, Tong and Li, Jundong and Huang, Zi},
  journal={IEEE Transactions on Knowledge and Data Engineering},
  volume={36},
  number={1},
  pages={335--355},
  year={2023},
  publisher={IEEE}
}

@inproceedings{ren2024representation,
  title={Representation learning with large language models for recommendation},
  author={Ren, Xubin and Wei, Wei and Xia, Lianghao and Su, Lixin and Cheng, Suqi and Wang, Junfeng and Yin, Dawei and Huang, Chao},
  booktitle={Proceedings of the ACM Web Conference 2024},
  pages={3464--3475},
  year={2024}
}

@inproceedings{cui2025comprehending,
  title={Comprehending knowledge graphs with large language models for recommender systems},
  author={Cui, Ziqiang and Weng, Yunpeng and Tang, Xing and Lyu, Fuyuan and Liu, Dugang and He, Xiuqiang and Ma, Chen},
  booktitle={Proceedings of the 48th International ACM SIGIR Conference on Research and Development in Information Retrieval},
  pages={1229--1239},
  year={2025}
}

@inproceedings{li2023alpt,
  title={Adaptive low-precision training for embeddings in click-through rate prediction},
  author={Li, Shiwei and Guo, Huifeng and Hou, Lu and Zhang, Wei and Tang, Xing and Tang, Ruiming and Zhang, Rui and Li, Ruixuan},
  booktitle={Proceedings of the AAAI Conference on Artificial Intelligence},
  volume={37},
  number={4},
  pages={4435--4443},
  year={2023}
}

@article{li2024optfp,
  title={Mixed-precision embeddings for large-scale recommendation models},
  author={Li, Shiwei and Hu, Zhuoqi and Tang, Xing and Wang, Haozhao and Xu, Shijie and Luo, Weihong and Li, Yuhua and He, Xiuqiang and Li, Ruixuan},
  journal={arXiv preprint arXiv:2409.20305},
  year={2024}
}


\newpage
\section*{NeurIPS Paper Checklist}

\begin{enumerate}

\item {\bf Claims}
    \item[] Question: Do the main claims made in the abstract and introduction accurately reflect the paper's contributions and scope?
    \item[] Answer: \answerYes{} 
    \item[] Justification: The abstract and introduction provide a concise yet comprehensive overview of our core innovation, key methodological contributions, and the specific research problem addressed in this work.
    \item[] Guidelines:
    \begin{itemize}
        \item The answer NA means that the abstract and introduction do not include the claims made in the paper.
        \item The abstract and/or introduction should clearly state the claims made, including the contributions made in the paper and important assumptions and limitations. A No or NA answer to this question will not be perceived well by the reviewers. 
        \item The claims made should match theoretical and experimental results, and reflect how much the results can be expected to generalize to other settings. 
        \item It is fine to include aspirational goals as motivation as long as it is clear that these goals are not attained by the paper. 
    \end{itemize}

\item {\bf Limitations}
    \item[] Question: Does the paper discuss the limitations of the work performed by the authors?
    \item[] Answer: \answerYes{} 
    \item[] Justification: We discuss the limitations of the work in the appendix.
    \item[] Guidelines:
    \begin{itemize}
        \item The answer NA means that the paper has no limitation while the answer No means that the paper has limitations, but those are not discussed in the paper. 
        \item The authors are encouraged to create a separate "Limitations" section in their paper.
        \item The paper should point out any strong assumptions and how robust the results are to violations of these assumptions (e.g., independence assumptions, noiseless settings, model well-specification, asymptotic approximations only holding locally). The authors should reflect on how these assumptions might be violated in practice and what the implications would be.
        \item The authors should reflect on the scope of the claims made, e.g., if the approach was only tested on a few datasets or with a few runs. In general, empirical results often depend on implicit assumptions, which should be articulated.
        \item The authors should reflect on the factors that influence the performance of the approach. For example, a facial recognition algorithm may perform poorly when image resolution is low or images are taken in low lighting. Or a speech-to-text system might not be used reliably to provide closed captions for online lectures because it fails to handle technical jargon.
        \item The authors should discuss the computational efficiency of the proposed algorithms and how they scale with dataset size.
        \item If applicable, the authors should discuss possible limitations of their approach to address problems of privacy and fairness.
        \item While the authors might fear that complete honesty about limitations might be used by reviewers as grounds for rejection, a worse outcome might be that reviewers discover limitations that aren't acknowledged in the paper. The authors should use their best judgment and recognize that individual actions in favor of transparency play an important role in developing norms that preserve the integrity of the community. Reviewers will be specifically instructed to not penalize honesty concerning limitations.
    \end{itemize}

\item {\bf Theory assumptions and proofs}
    \item[] Question: For each theoretical result, does the paper provide the full set of assumptions and a complete (and correct) proof?
    \item[] Answer: \answerNA{} 
    \item[] Justification: This paper does not include theoretical results.
    \item[] Guidelines:
    \begin{itemize}
        \item The answer NA means that the paper does not include theoretical results. 
        \item All the theorems, formulas, and proofs in the paper should be numbered and cross-referenced.
        \item All assumptions should be clearly stated or referenced in the statement of any theorems.
        \item The proofs can either appear in the main paper or the supplemental material, but if they appear in the supplemental material, the authors are encouraged to provide a short proof sketch to provide intuition. 
        \item Inversely, any informal proof provided in the core of the paper should be complemented by formal proofs provided in appendix or supplemental material.
        \item Theorems and Lemmas that the proof relies upon should be properly referenced. 
    \end{itemize}

    \item {\bf Experimental result reproducibility}
    \item[] Question: Does the paper fully disclose all the information needed to reproduce the main experimental results of the paper to the extent that it affects the main claims and/or conclusions of the paper (regardless of whether the code and data are provided or not)?
    \item[] Answer: \answerYes{} 
    \item[] Justification: This paper provides full disclosure of all information necessary to reproduce our key experimental results, and provides source code on an anonymized GitHub repository.
    \item[] Guidelines:
    \begin{itemize}
        \item The answer NA means that the paper does not include experiments.
        \item If the paper includes experiments, a No answer to this question will not be perceived well by the reviewers: Making the paper reproducible is important, regardless of whether the code and data are provided or not.
        \item If the contribution is a dataset and/or model, the authors should describe the steps taken to make their results reproducible or verifiable. 
        \item Depending on the contribution, reproducibility can be accomplished in various ways. For example, if the contribution is a novel architecture, describing the architecture fully might suffice, or if the contribution is a specific model and empirical evaluation, it may be necessary to either make it possible for others to replicate the model with the same dataset, or provide access to the model. In general. releasing code and data is often one good way to accomplish this, but reproducibility can also be provided via detailed instructions for how to replicate the results, access to a hosted model (e.g., in the case of a large language model), releasing of a model checkpoint, or other means that are appropriate to the research performed.
        \item While NeurIPS does not require releasing code, the conference does require all submissions to provide some reasonable avenue for reproducibility, which may depend on the nature of the contribution. For example
        \begin{enumerate}
            \item If the contribution is primarily a new algorithm, the paper should make it clear how to reproduce that algorithm.
            \item If the contribution is primarily a new model architecture, the paper should describe the architecture clearly and fully.
            \item If the contribution is a new model (e.g., a large language model), then there should either be a way to access this model for reproducing the results or a way to reproduce the model (e.g., with an open-source dataset or instructions for how to construct the dataset).
            \item We recognize that reproducibility may be tricky in some cases, in which case authors are welcome to describe the particular way they provide for reproducibility. In the case of closed-source models, it may be that access to the model is limited in some way (e.g., to registered users), but it should be possible for other researchers to have some path to reproducing or verifying the results.
        \end{enumerate}
    \end{itemize}

\item {\bf Open access to data and code}
    \item[] Question: Does the paper provide open access to the data and code, with sufficient instructions to faithfully reproduce the main experimental results, as described in supplemental material?
    \item[] Answer: \answerYes{} 
    \item[] Justification: The four datasets used in this paper are all public, and we offer their links. We also provide our source code on the anonymized GitHub repository.
    \item[] Guidelines:
    \begin{itemize}
        \item The answer NA means that paper does not include experiments requiring code.
        \item Please see the NeurIPS code and data submission guidelines (\url{https://nips.cc/public/guides/CodeSubmissionPolicy}) for more details.
        \item While we encourage the release of code and data, we understand that this might not be possible, so “No” is an acceptable answer. Papers cannot be rejected simply for not including code, unless this is central to the contribution (e.g., for a new open-source benchmark).
        \item The instructions should contain the exact command and environment needed to run to reproduce the results. See the NeurIPS code and data submission guidelines (\url{https://nips.cc/public/guides/CodeSubmissionPolicy}) for more details.
        \item The authors should provide instructions on data access and preparation, including how to access the raw data, preprocessed data, intermediate data, and generated data, etc.
        \item The authors should provide scripts to reproduce all experimental results for the new proposed method and baselines. If only a subset of experiments are reproducible, they should state which ones are omitted from the script and why.
        \item At submission time, to preserve anonymity, the authors should release anonymized versions (if applicable).
        \item Providing as much information as possible in supplemental material (appended to the paper) is recommended, but including URLs to data and code is permitted.
    \end{itemize}

\item {\bf Experimental setting/details}
    \item[] Question: Does the paper specify all the training and test details (e.g., data splits, hyperparameters, how they were chosen, type of optimizer, etc.) necessary to understand the results?
    \item[] Answer: \answerYes{} 
    \item[] Justification: We specify all training and test details, including data splits, hyperparameters, how they were chosen, type of optimizer, etc. necessary to understand the results in the experiment section and the appendix.
    \item[] Guidelines:
    \begin{itemize}
        \item The answer NA means that the paper does not include experiments.
        \item The experimental setting should be presented in the core of the paper to a level of detail that is necessary to appreciate the results and make sense of them.
        \item The full details can be provided either with the code, in appendix, or as supplemental material.
    \end{itemize}

\item {\bf Experiment statistical significance}
    \item[] Question: Does the paper report error bars suitably and correctly defined or other appropriate information about the statistical significance of the experiments?
    \item[] Answer: \answerYes{} 
    \item[] Justification: The experimental results in this paper are confirmed by a paired t-test with a significance level of 0.01.
    \item[] Guidelines:
    \begin{itemize}
        \item The answer NA means that the paper does not include experiments.
        \item The authors should answer "Yes" if the results are accompanied by error bars, confidence intervals, or statistical significance tests, at least for the experiments that support the main claims of the paper.
        \item The factors of variability that the error bars are capturing should be clearly stated (for example, train/test split, initialization, random drawing of some parameter, or overall run with given experimental conditions).
        \item The method for calculating the error bars should be explained (closed form formula, call to a library function, bootstrap, etc.)
        \item The assumptions made should be given (e.g., Normally distributed errors).
        \item It should be clear whether the error bar is the standard deviation or the standard error of the mean.
        \item It is OK to report 1-sigma error bars, but one should state it. The authors should preferably report a 2-sigma error bar than state that they have a 96\% CI, if the hypothesis of Normality of errors is not verified.
        \item For asymmetric distributions, the authors should be careful not to show in tables or figures symmetric error bars that would yield results that are out of range (e.g. negative error rates).
        \item If error bars are reported in tables or plots, The authors should explain in the text how they were calculated and reference the corresponding figures or tables in the text.
    \end{itemize}

\item {\bf Experiments compute resources}
    \item[] Question: For each experiment, does the paper provide sufficient information on the computer resources (type of compute workers, memory, time of execution) needed to reproduce the experiments?
    \item[] Answer: \answerYes{} 
    \item[] Justification: The paper provides sufficient information on the computer resources needed to reproduce the experiments.
    \item[] Guidelines:
    \begin{itemize}
        \item The answer NA means that the paper does not include experiments.
        \item The paper should indicate the type of compute workers CPU or GPU, internal cluster, or cloud provider, including relevant memory and storage.
        \item The paper should provide the amount of compute required for each of the individual experimental runs as well as estimate the total compute. 
        \item The paper should disclose whether the full research project required more compute than the experiments reported in the paper (e.g., preliminary or failed experiments that didn't make it into the paper). 
    \end{itemize}
    
\item {\bf Code of ethics}
    \item[] Question: Does the research conducted in the paper conform, in every respect, with the NeurIPS Code of Ethics \url{https://neurips.cc/public/EthicsGuidelines}?
    \item[] Answer: \answerYes{} 
    \item[] Justification: The research conducted in the paper conforms, in every respect, with the NeurIPS Code of Ethics.
    \item[] Guidelines:
    \begin{itemize}
        \item The answer NA means that the authors have not reviewed the NeurIPS Code of Ethics.
        \item If the authors answer No, they should explain the special circumstances that require a deviation from the Code of Ethics.
        \item The authors should make sure to preserve anonymity (e.g., if there is a special consideration due to laws or regulations in their jurisdiction).
    \end{itemize}

\item {\bf Broader impacts}
    \item[] Question: Does the paper discuss both potential positive societal impacts and negative societal impacts of the work performed?
    \item[] Answer: \answerYes{} 
    \item[] Justification: The paper discussed both potential positive societal impacts and negative societal impacts in the appendix.
    \item[] Guidelines:
    \begin{itemize}
        \item The answer NA means that there is no societal impact of the work performed.
        \item If the authors answer NA or No, they should explain why their work has no societal impact or why the paper does not address societal impact.
        \item Examples of negative societal impacts include potential malicious or unintended uses (e.g., disinformation, generating fake profiles, surveillance), fairness considerations (e.g., deployment of technologies that could make decisions that unfairly impact specific groups), privacy considerations, and security considerations.
        \item The conference expects that many papers will be foundational research and not tied to particular applications, let alone deployments. However, if there is a direct path to any negative applications, the authors should point it out. For example, it is legitimate to point out that an improvement in the quality of generative models could be used to generate deepfakes for disinformation. On the other hand, it is not needed to point out that a generic algorithm for optimizing neural networks could enable people to train models that generate Deepfakes faster.
        \item The authors should consider possible harms that could arise when the technology is being used as intended and functioning correctly, harms that could arise when the technology is being used as intended but gives incorrect results, and harms following from (intentional or unintentional) misuse of the technology.
        \item If there are negative societal impacts, the authors could also discuss possible mitigation strategies (e.g., gated release of models, providing defenses in addition to attacks, mechanisms for monitoring misuse, mechanisms to monitor how a system learns from feedback over time, improving the efficiency and accessibility of ML).
    \end{itemize}
    
\item {\bf Safeguards}
    \item[] Question: Does the paper describe safeguards that have been put in place for responsible release of data or models that have a high risk for misuse (e.g., pretrained language models, image generators, or scraped datasets)?
    \item[] Answer: \answerNA{} 
    \item[] Justification: There is no risk of misuse of the proposed method and the datasets used in the paper are open-sourced.
    \item[] Guidelines:
    \begin{itemize}
        \item The answer NA means that the paper poses no such risks.
        \item Released models that have a high risk for misuse or dual-use should be released with necessary safeguards to allow for controlled use of the model, for example by requiring that users adhere to usage guidelines or restrictions to access the model or implementing safety filters. 
        \item Datasets that have been scraped from the Internet could pose safety risks. The authors should describe how they avoided releasing unsafe images.
        \item We recognize that providing effective safeguards is challenging, and many papers do not require this, but we encourage authors to take this into account and make a best faith effort.
    \end{itemize}

\item {\bf Licenses for existing assets}
    \item[] Question: Are the creators or original owners of assets (e.g., code, data, models), used in the paper, properly credited and are the license and terms of use explicitly mentioned and properly respected?
    \item[] Answer: \answerYes{} 
    \item[] Justification: All third-party assets are properly credited through citations and in-text acknowledgments.
    \item[] Guidelines:
    \begin{itemize}
        \item The answer NA means that the paper does not use existing assets.
        \item The authors should cite the original paper that produced the code package or dataset.
        \item The authors should state which version of the asset is used and, if possible, include a URL.
        \item The name of the license (e.g., CC-BY 4.0) should be included for each asset.
        \item For scraped data from a particular source (e.g., website), the copyright and terms of service of that source should be provided.
        \item If assets are released, the license, copyright information, and terms of use in the package should be provided. For popular datasets, \url{paperswithcode.com/datasets} has curated licenses for some datasets. Their licensing guide can help determine the license of a dataset.
        \item For existing datasets that are re-packaged, both the original license and the license of the derived asset (if it has changed) should be provided.
        \item If this information is not available online, the authors are encouraged to reach out to the asset's creators.
    \end{itemize}

\item {\bf New assets}
    \item[] Question: Are new assets introduced in the paper well documented and is the documentation provided alongside the assets?
    \item[] Answer: \answerYes{} 
    \item[] Justification: We communicate the details of the dataset/code/model as part of our submission with an anonymized github URL.
    \item[] Guidelines:
    \begin{itemize}
        \item The answer NA means that the paper does not release new assets.
        \item Researchers should communicate the details of the dataset/code/model as part of their submissions via structured templates. This includes details about training, license, limitations, etc. 
        \item The paper should discuss whether and how consent was obtained from people whose asset is used.
        \item At submission time, remember to anonymize your assets (if applicable). You can either create an anonymized URL or include an anonymized zip file.
    \end{itemize}

\item {\bf Crowdsourcing and research with human subjects}
    \item[] Question: For crowdsourcing experiments and research with human subjects, does the paper include the full text of instructions given to participants and screenshots, if applicable, as well as details about compensation (if any)? 
    \item[] Answer: \answerNA{} 
    \item[] Justification: The paper does not involve crowdsourcing nor research with human subjects.
    \item[] Guidelines:
    \begin{itemize}
        \item The answer NA means that the paper does not involve crowdsourcing nor research with human subjects.
        \item Including this information in the supplemental material is fine, but if the main contribution of the paper involves human subjects, then as much detail as possible should be included in the main paper. 
        \item According to the NeurIPS Code of Ethics, workers involved in data collection, curation, or other labor should be paid at least the minimum wage in the country of the data collector. 
    \end{itemize}

\item {\bf Institutional review board (IRB) approvals or equivalent for research with human subjects}
    \item[] Question: Does the paper describe potential risks incurred by study participants, whether such risks were disclosed to the subjects, and whether Institutional Review Board (IRB) approvals (or an equivalent approval/review based on the requirements of your country or institution) were obtained?
    \item[] Answer: \answerNA{} 
    \item[] Justification: The paper does not involve crowdsourcing nor research with human subjects.
    \item[] Guidelines:
    \begin{itemize}
        \item The answer NA means that the paper does not involve crowdsourcing nor research with human subjects.
        \item Depending on the country in which research is conducted, IRB approval (or equivalent) may be required for any human subjects research. If you obtained IRB approval, you should clearly state this in the paper. 
        \item We recognize that the procedures for this may vary significantly between institutions and locations, and we expect authors to adhere to the NeurIPS Code of Ethics and the guidelines for their institution. 
        \item For initial submissions, do not include any information that would break anonymity (if applicable), such as the institution conducting the review.
    \end{itemize}

\item {\bf Declaration of LLM usage}
    \item[] Question: Does the paper describe the usage of LLMs if it is an important, original, or non-standard component of the core methods in this research? Note that if the LLM is used only for writing, editing, or formatting purposes and does not impact the core methodology, scientific rigorousness, or originality of the research, declaration is not required.
    \item[] Answer: \answerNA{} 
    \item[] Justification: The core method development in this research does not involve LLMs as any important, original, or non-standard components.
    \item[] Guidelines:
    \begin{itemize}
        \item The answer NA means that the core method development in this research does not involve LLMs as any important, original, or non-standard components.
        \item Please refer to our LLM policy (\url{https://neurips.cc/Conferences/2025/LLM}) for what should or should not be described.
    \end{itemize}

\end{enumerate}

\newpage
\appendix

\section{Technical Supplement to SRA-CL}
\subsection{Prompt Template}
\begin{figure}[h]
  \centering
  \includegraphics[width=0.98\columnwidth]{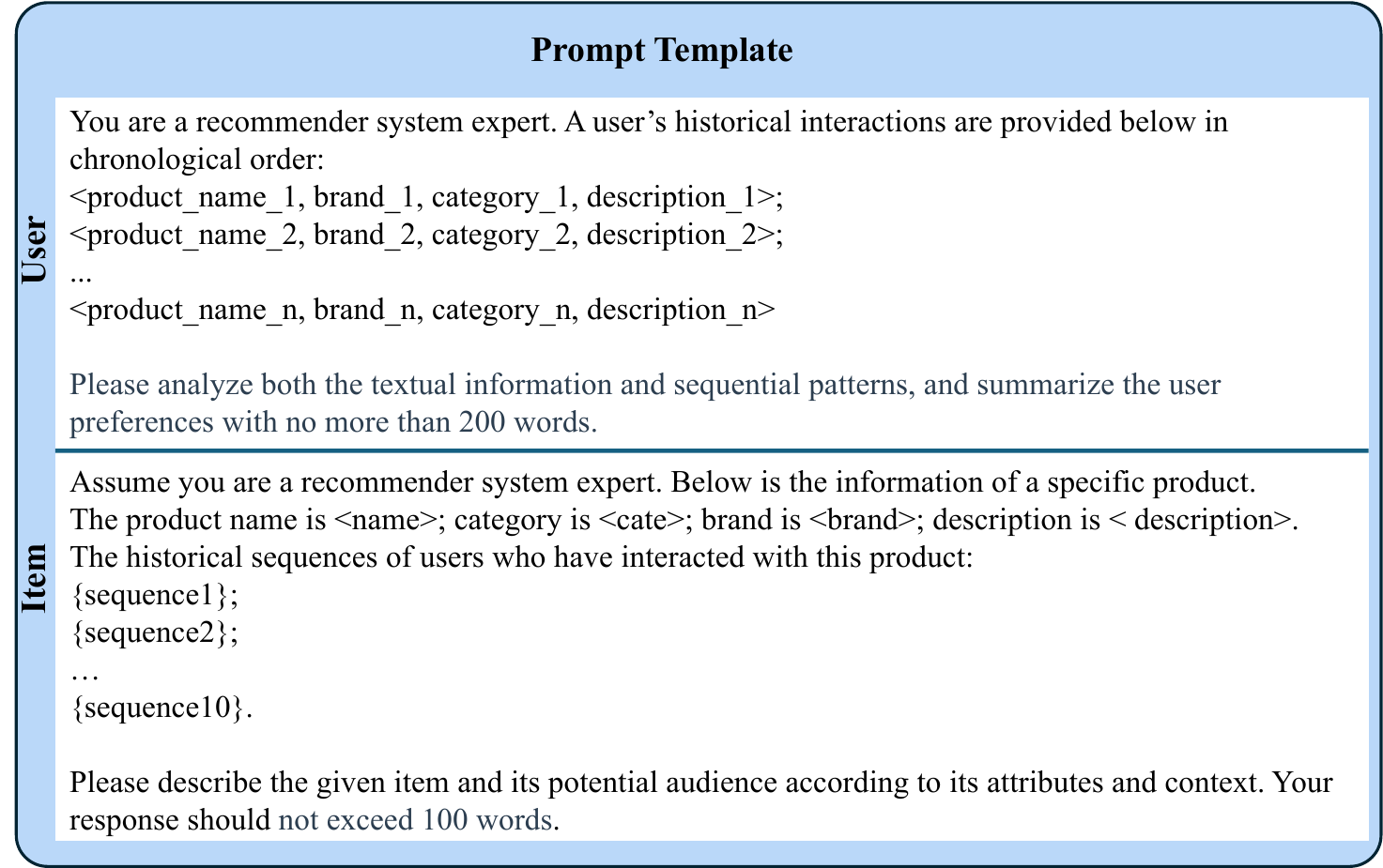}
  \caption{Prompt Template.} 
   \label{prompt}
\end{figure}

In this section, we provide a detailed description of the LLM prompt templates employed in our study. Specifically, to enhance the model's ability to comprehend user preferences and items, we have meticulously designed specialized prompts, as illustrated in Figure \ref{prompt}.

\subsection{Efficiency Analysis} 
\textbf{Inference Efficiency}. During inference, only the recommendation backbone is utilized. The contrastive learning tasks and the LLMs’ semantic embeddings are not involved in the inference process. This ensures that our framework can be deployed in real-world applications without incurring any additional inference latency from incorporating LLMs.

\textbf{Training Efficiency}. The training process of our method consists of two stages: In the first stage, we use an LLM API to obtain semantic information and convert it into embeddings, which are then cached to construct contrastive sample indices. The primary time cost in this stage comes from the API calls. However, by employing asynchronous concurrency, this step can be completed within a few hours. Crucially, this stage is performed once and requires no repetition during model training. In the second stage, we use the pre-constructed contrastive sample index to train the recommendation model. Regarding the computational complexity of this stage, our method maintains comparable time complexity to general ID-based contrastive recommendation approaches. The only additional overhead during training compared to conventional contrastive recommendation models comes from the lightweight learnable sample synthesis module whose parameter size is negligible compared to that of the main recommendation model.

\section{Experimental Setting Details}

\begin{table}[t] 
\renewcommand\arraystretch{1.0}
	\centering
	\caption{Dataset statistics.}  
 \scalebox{1.0}{
\begin{tabular}{lrrrrr}
\toprule
Datasets & \#Users  & \#Items  & \#Actions & Avg. Length & Density  \\
\midrule
Yelp &19,936 & 14,587 &207,952 &10.4 &0.07\%  \\
Sports &35,598&   18,357 & 296,337 &8.3  &0.05\%    \\
Beauty &22,363 &12,101  &198,502  &8.8  &0.07\%   \\
Office &4,905&   2,420 & 53,258 &10.9  &0.45\%    \\
\bottomrule
\end{tabular}}
\label{dataset}
\end{table}

\subsection{\textbf{Datasets}} \label{appendix:datasets}
We conducted experiments on four public real-world datasets: Yelp, Sports, Beauty, and Office. The statistics for these datasets are presented in Table \ref{dataset}. These datasets cover a diverse range of application scenarios. The Yelp dataset\footnote{https://www.yelp.com/dataset} is widely used for business recommendations. The Sports, Beauty, and Office datasets are sourced from Amazon\footnote{http://jmcauley.ucsd.edu/data/amazon/}, one of the largest e-commerce platforms. Following previous studies \cite{liu2021contrastive,xie2022contrastive,qin2023meta}, the
users and items that have fewer than five interactions are removed.

\subsection{\textbf{Baseline Methods}}
To ensure a comprehensive assessment, we compare our method with 12 baseline methods, categorized into three groups: classical methods (GRU4Rec, SASRec, BERT4Rec), contrastive learning-based methods ($\rm {{S}^{3}\text{-}{Rec}}$, CL4SRec, CoSeRec, ICLRec, DuoRec, MCLRec, ICSRec), and LLM-enhanced methods (LRD, LLM-ESR).

\begin{itemize}[leftmargin=1em]
\item \textbf{GRU4Rec} \cite{hidasi2015session} applies recurrent neural networks (RNN) to sequential recommendation.
\item \textbf{SASRec} \cite{kang2018self} is the first work to utilize the self-attention mechanism for sequential recommendation.
\item \textbf{BERT4Rec} \cite{sun2019bert4rec} employs the BERT \cite{devlin2018bert} framework to capture the context information of user behaviors.
\item $\rm {\textbf{S}^{3}\text{-}\textbf{Rec}}$ \cite{zhou2020s3} leverages four self-supervised objectives to uncover the inherent correlations within the data.
\item \textbf{CL4SRec} \cite{xie2022contrastive} proposes three random augmentation operators to generate positive samples for contrastive learning.
\item $\textbf{CoSeRec}$ \cite{liu2021contrastive} introduces two additional informative augmentation operators, building upon the foundation of CL4SRec. 
\item $\textbf{ICLRec}$ \cite{chen2022intent} clusters user interests into distinct categories and brings the representations of users with similar interests closer together.
\item \textbf{DuoRec} \cite{qiu2022contrastive} combines a model-level dropout augmentation and a sampling strategy for choosing hard positive samples.
\item \textbf{MCLRec} \cite{qin2023meta} integrates CL4SRec's random data augmentation for the input sequence and employs MLP layers for model-level augmentation.
\item \textbf{ICSRec} \cite{qin2024intent} is an improvement on ICLRec, further segmenting a user's sequential behaviors into multiple subsequences to generate finer-grained user intentions for contrastive learning.
\item \textbf{LRD} \cite{yang2024sequential} is an LLM-based method. It leverages LLMs to discover new relations between items and reconstructs one item based on the relation and another item.
\item \textbf{RLMRec} \cite{ren2024representation} utilizes LLMs to generate text profiles and combine their semantic embeddings with recommendation models.
\item \textbf{LLM-ESR} \cite{liu2024llm} is also an LLM-based method. It addresses the long-tail problem by simultaneously leveraging collaborative signals and semantic information through the dual-view modeling and self-distillation.
\end{itemize}

\subsection{Implementation Details} \label{appendix:imple details}
All experiments are conducted with a single V100 GPU. The embedding size for all methods is set to 64 for a fair comparison. 
We use a training batch size of 256 and employ the Adam optimizer with a learning rate of 0.001. The dropout rate is set to 0.5 for both the embedding layer and the hidden layers across all datasets. Following previous studies \cite{yang2024sequential}, we set the maximum sequence length to 20 for all datasets. The early stopping is applied if the metrics on the validation set do not improve over 10 consecutive epochs. 
Our method is model-agnostic and can be applied to any sequential recommendation model. The transformer backbone mentioned in Sec. \ref{backbone} comprises two layers, each with two attention heads. 
For the LLM, we select DeepSeek-V3, a robust large language model that demonstrates exceptional performance on both standard benchmarks and open-ended generation evaluations. 
For detailed information about DeepSeek, please refer to their official website\footnote{https://github.com/deepseek-ai/DeepSeek-V3}. Specifically, we utilize DeepSeek-V3 by invoking its API\footnote{https://api-docs.deepseek.com/}. To reduce text randomness of the LLM, we set the temperature $\tau$ to 0 and the top-$p$ to 0.001. For the text embedding model $\mathcal{M}$, we use the pre-trained SimCSE-RoBERTa\footnote{https://huggingface.co/princeton-nlp/sup-simcse-roberta-large} from Hugging Face. Identical settings are used for baselines that involve LLMs and text embeddings to ensure fairness.

\section{Additional Results \& Analysis} \label{appendix:additional results}

 


\subsection{Discussion on Learnable Sample Synthesis} \label{Discussion on Learnable Sample Synthesis}

\textbf{Inter-User Contrastive Learning}. User preferences exhibit significant heterogeneity across individuals. Sole reliance on hard rules, such as selecting a user from the current user's dedicated candidate pool as the positive sample, may yield suboptimal solutions. Our experiments (as shown in Table \ref{ablation study} ``w/o learn.'') validated this. To enhance contrastive sample construction, we introduce a learnable sample synthesizer that optimizes the contrastive sample generation process during model training for inter-user contrastive learning.

\textbf{Intra-User Contrastive Learning}. Our preliminary experiments also explored the use of learnable synthesizers (analogous to inter-contrastive learning approaches) for generating substitute items, yet yielded no measurable performance improvements (shown in Table \ref{item learnable results}). Our analysis suggests this results from the inherent nature of item semantics being more readily interpretable and quantifiable than user preferences. Therefore, directly identifying appropriate substitutes from semantically similar candidate pools is simpler and more reliable compared to matching users with analogous preference patterns.
\begin{table*}[h]  
	\centering
\caption{Performance impact of learnable versus non-learnable sample synthesis strategies in \textbf{intra-user} contrastive learning.}
\label{item learnable results}
\setlength\tabcolsep{3pt}
 \scalebox{0.9}{
	\begin{tabular}{l|cc|cc|cc|cc}
		\toprule
		\multirow{2}*{} 
            & \multicolumn{2}{c|}{Yelp} 
            &\multicolumn{2}{c|}{Sports} 
            &\multicolumn{2}{c|}{Beauty} 
            &\multicolumn{2}{c}{Office}
            \\ 
    \cmidrule(lr){2-9} 
 
& HR@20 & NDCG@20 
& HR@20 & NDCG@20 
& HR@20 & NDCG@20  
& HR@20 & NDCG@20 
\\

\midrule 
Learnable  &{0.1276} &{0.0531} &{0.0825} &{0.0344} &{0.1309} &{0.0561} &{0.1706} &{0.0722}
\\
Unlearnable &{0.1282} &{0.0533} &{0.0823} &{0.0347} &{0.1314} &{0.0568} &{0.1702} &{0.0725}
\\
\bottomrule
\end{tabular} }
\end{table*}

\subsection{Additional Comparison Results} \label{Additional Comparison Results}
\begin{table*}[t]  
\renewcommand\arraystretch{1.1}
	\centering
\caption{Additional comparison results for HR@10 and NDCG@10. Bold font indicates the best performance, while underlined values represent the second-best. ``ND'' represents for ``NDCG''. Our method SRA-CL achieves state-of-the-art results among all methods, as confirmed by a paired t-test with a significance level of 0.01.}
\label{comparison results hr10}
\setlength\tabcolsep{4pt}
 \scalebox{0.9}{
	\begin{tabular}{l|cc|cc|cc|cc}
		\toprule
		\multirow{2}*{Model} 
            & \multicolumn{2}{c|}{Yelp} 
            &\multicolumn{2}{c|}{Sports} 
            &\multicolumn{2}{c|}{Beauty} 
            &\multicolumn{2}{c}{Office}
            \\ 
    \cmidrule(lr){2-9} 
 
& HR@10 & ND@10 
& HR@10 & ND@10 
& HR@10 & ND@10 
& HR@10 & ND@10 
\\

\midrule 
GRU4Rec  &0.0362 &0.0173 &0.0193 &0.0096 &0.0279 &0.0137 &0.0540 &0.0260
\\
SASRec   &0.0572 &0.0308 &0.0304 &0.0157 &0.0612 &0.0336 &0.0791 &0.0348
\\
BERT4Rec  &0.0582 &0.0311 &0.0349 &0.0189 &0.0628 &0.0352 &0.0821 &0.0376
\\
\midrule
$\mathrm {\text{S}^3\text{-}\text{Rec}}$ &0.0612 &0.0339 &0.0385 &0.0204 &0.0647 &0.0327 &0.0931 &0.0426
\\
CL4SRec &0.0583 &0.0315 &0.0358 &0.0189 &0.0649 &0.0329 &0.0695 &0.0322
\\
CoSeRec  &0.0607 &0.0309 &0.0439 &0.0244 &0.0725 &0.0410 &0.0782 &0.0412
\\
ICLRec  &0.0598 &0.0328 &0.0428 &0.0235 &0.0713 &0.0396 &0.0922 &0.0411
\\
DuoRec  &0.0747 &0.0380 &0.0474 &0.0242 &0.0841 &0.0443 &0.1015 &0.0519 
\\
MCLRec &0.0721 &0.0378 &0.0498 &0.0257 &0.0870 &0.0442 &0.1036 &0.0538 
\\
ICSRec &0.0738 &0.0380 &0.0487 &0.0243 &0.0844 &0.0437 &0.1034 &0.0540 
\\
\midrule
LRD &0.0693 &0.0357 &0.0376 &0.0191 &0.0620 &0.0294 &0.0887 &0.0431
\\
RLMRec &0.0709 &0.0371 &0.0426 &0.0238 &0.0764 &0.0439 &0.0927 &0.0496
\\
LLM-ESR &0.0669 &0.0353 &0.0415 &0.0221 &0.0750 &0.0435 &0.0889 &0.0468
\\
\midrule
\textbf{SRA-CL} &\textbf{0.0817} &\textbf{0.0419} &\textbf{0.0539} &\textbf{0.0274} &\textbf{0.0924} &\textbf{0.0469} &\textbf{0.1111} &\textbf{0.0575}
\\
\midrule
Improvement &9.37\% &10.26\% &8.23\% &6.61\% &6.21\% &6.11\% &7.24\% &6.48\%
\\
\bottomrule
\end{tabular} }
\end{table*}
We provide additional comparison results (HR@10 and NDCG@10) of different methods in Table \ref{comparison results hr10}. The experimental results demonstrate that our method outperforms all baselines across all datasets, further validating its superiority.

\subsection{Additional Results for Hyperparameter Experiments}
\begin{figure}[t]
\setlength{\abovecaptionskip}{-0.1mm} 
  \centering
  \includegraphics[width=0.8\columnwidth]{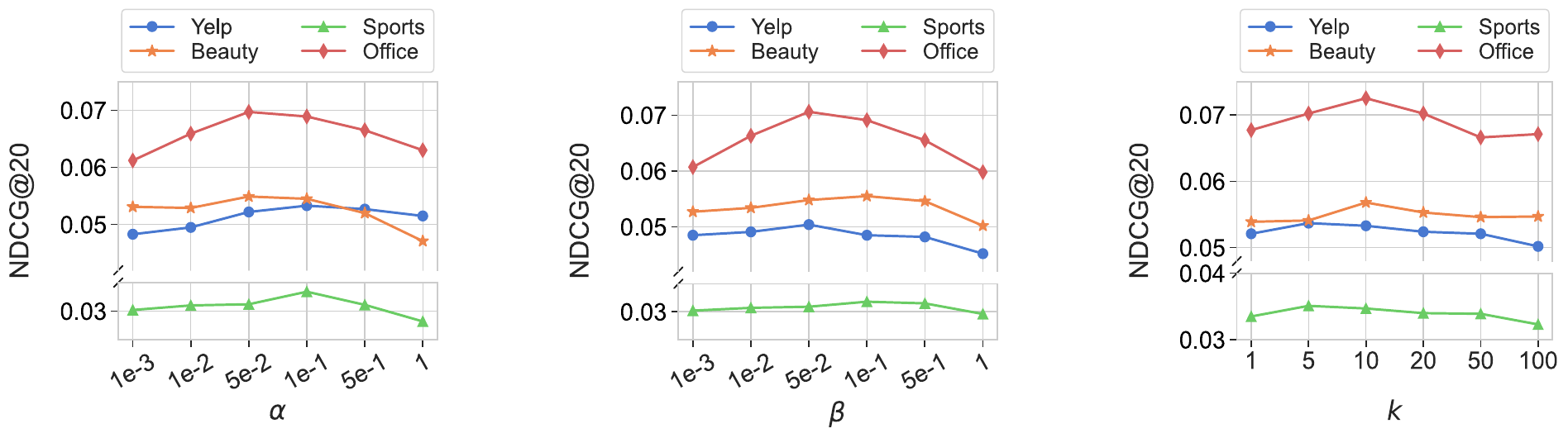}
  \caption{Hyperparameter experiments on the weight of $\mathcal{L}_{CS}$ ($\alpha$), the weight of $\mathcal{L}_{IS}$ ($\beta$), and the number of retrieved users/items ($k$) (NDCG results).} 
   \label{param_study_loss_ndcg}
\end{figure}

Due to space constraints, we only present HR@20 in Figure \ref{param_study_loss} of the main text for hyperparameter study. Here, we additionally report the NDCG@20 evaluation results in Figure \ref{param_study_loss_ndcg}, providing complementary performance metrics for comprehensive analysis. As shown, the trend in NDCG@20 closely aligns with that of HR@20.

\section{Other Discussions}
\subsection{Limitation}
Considering computational budgets and resource limitations, we specifically analyzed how two selected LLMs (DeepSeek and Qwen) affect our framework's effectiveness. While more LLMs might yield different results, our study focused on these representative models.

\subsection{Broader Impacts}
SRA-CL demonstrates significant improvements in sequential recommendation accuracy (positive impact), with potential applicability to real-world platforms. Like all recommendation systems, its personalized nature may occasionally limit content diversity, though this effect is inherent to the recommendation paradigm rather than unique to our method.

\end{document}